\documentclass{emulateapj}
\usepackage{topcapt}
\usepackage{booktabs}
\usepackage{natbib}
\usepackage{rotating}
\usepackage{longtable}
\shorttitle{IMF in NGC4214}
\shortauthors{Andrews et al.}

\begin{document}

\title{An IMF Study of the Dwarf Starburst Galaxy NGC 4214}\footnotetext[0]{Based on observations made with the NASA/ESA Hubble Space Telescope, obtained at the Space Telescope Science Institute, which is operated by the Association of Universities for Research in Astronomy, Inc., under NASA contract NAS 5-26555. These observations are associated with program \# GO-11360.}

\author{J.E. Andrews\altaffilmark{1}, D. Calzetti\altaffilmark{1}, R. Chandar\altaffilmark{2}, J.C. Lee\altaffilmark{3,4}, B.G. Elmegreen\altaffilmark{5}, R.C. Kennicutt\altaffilmark{6}, B. Whitmore\altaffilmark{3}, J.S. Kissel\altaffilmark{7}, Robert L. da Silva\altaffilmark{8,9}, Mark R. Krumholz\altaffilmark{8}, R.W. O'Connell\altaffilmark{10}, M.A. Dopita\altaffilmark{11,12}, Jay A. Frogel\altaffilmark{13}, Hwihyun Kim\altaffilmark{14}}

\altaffiltext{1}{Department of Astronomy, University of Massachusetts, Amherst, MA 01003, USA; jandrews@astro.umass.edu, callzetti@astro.umass.edu}
\altaffiltext{2}{Department of Physics and Astronomy, University of Toledo, Toledo, OH 43606, USA}
\altaffiltext{3}{Space Telescope Science Institute, 3700 San Martin Drive, Baltimore, MD 21218, USA}
\altaffiltext{4}{Visiting Astronomer, Spitzer Science Center, Caltech Pasadena, CA 91125 }
\altaffiltext{5}{IBM T.J. Watson Research Center, Yorktown Heights, NY, USA}
\altaffiltext{6}{Institute of Astronomy, Cambridge University, Cambridge, UK}
\altaffiltext{7}{Kavli Institute for Astrophysics and Space Research, Massachusetts Institute of Technology, Cambridge, MA 02139}
\altaffiltext{8}{Department of Astronomy and Astrophysics, University of California, 1156 High Street, Santa Cruz, CA 95064, USA}
\altaffiltext{9}{NSF Graduate Research Fellow}
\altaffiltext{10}{Department of Astronomy, University of Virginia, P.O. Box 3818, Charlottesville, VA, 22903, USA}
\altaffiltext{11}{Research School of Astronomy and Astrophysics, Australian National University, Cotter Rd., Weston ACT 2611, Australia}
\altaffiltext{12}{Astronomy Department, King Abdulaziz University, P.O. Box 80203, Jeddah, Saudi Arabia}
\altaffiltext{13}{Galaxies Unlimited, 1 Tremblant Court, Lutherville, MD, USA}
\altaffiltext{14}{School of Earth and Space Exploration, Arizona State University, Tempe, AZ 85287-1404, USA}

\begin{abstract}
The production rate of ionizing photons in young ($\leq$ 8 Myr), unresolved stellar clusters in the nearby irregular galaxy NGC 4214 is probed using multi-wavelength \emph{Hubble Space Telescope} WFC3 data.  We normalize the ionizing photon rate by the cluster mass to investigate the upper end of the stellar initial mass function (IMF).  We have found that within the uncertainties the upper end of the stellar IMF appears to be universal in this galaxy, and that deviations from a universal IMF can be attributed to stochastic sampling of stars in clusters with masses $\lessapprox$ 10$^{3}$ M$_{\sun}$.  Furthermore, we have found that there does not seem to be a dependence of the maximum stellar mass on the cluster mass. We have also found that for massive clusters, feedback may cause an underrepresentation in H$\alpha$ luminosities, which needs to be taken into account when conducting this type of analysis. 
\end{abstract}


\section{Introduction}

The stellar initial mass function (IMF), the distribution of stellar masses in newly formed stellar populations is essential for understanding the evolution and star formation histories of galaxies.  Whether it is universal or dependent on environment has been a highly contested issue over the past few years.  While IMF measurements in high density environments like the Milky Way and Magellanic Clouds have indicated an invariant IMF \citep{2010ARA&A..48..339B,2003ARA&A..41...15M,2011ApJ...739L..46O}, other claims of a non-universal IMF have been made \citep{2011ApJ...735L..13V,2008MNRAS.391..363W,2007MNRAS.379..985F,2011arXiv1112.3340K,2012Natur.484..485C}.   In particular, star-forming dwarf galaxies may show a deficiency in the ionizing photon rate per unit UV or optical luminosity \citep{2008ApJ...675..163H,2009ApJ...706..599L,2009ApJ...695..765M,2009ApJ...706.1527B,2011MNRAS.415.1647G}. However, \citet{2011ApJ...741L..26F}  and \citet{2012ApJ...744...44W} have shown that stochasticity in populating the IMF or bursts of star formation can explain the observed variations in L$_{H\alpha}$/L$_{FUV}$ so that an unusual IMF is not required. Clearly this is an unresolved issue that is essential to understanding the fundamental evolution of galaxies.

Traditional methods for IMF measurements in the Milky Way, Large Magellanic Cloud (LMC), and Small Magellanic Cloud (SMC) are to count individual stars in clusters $\leq$ 3-5 Myr old that still retain their most massive stars \citep[for example]{2009ApJ...697L..58A,2008AJ....135..173S,2000ApJ...533..203S}. Generally only 20$\%$ of stellar clusters will survive early mass loss to live longer than 10 Myr \citep[``infant mortality'']{2003ARA&A..41...57L}, so catching them very early is essential for observing the full stellar population. Even nearby, significant crowding can cause these star counts to be incomplete, with high mass stars and low mass stars suffering from different selection biases: low mass stars are generally harder to count, due to the inability to easily detect smaller, fainter stars and due to dynamical ejection of the low mass stars, while massive stars may suffer from confusion due to their sinking towards the center of the cluster \citep{2009A&A...495..147A,2008ApJ...677.1278M}.  This method becomes progressively less effective at distances outside of the Magellanic Clouds ($\sim$ 50 kpc), even with the resolution of the \emph{Hubble Space Telescope} (HST).

 \begin{figure*}[t!] 
   \centering
   \includegraphics[width=6.3in]{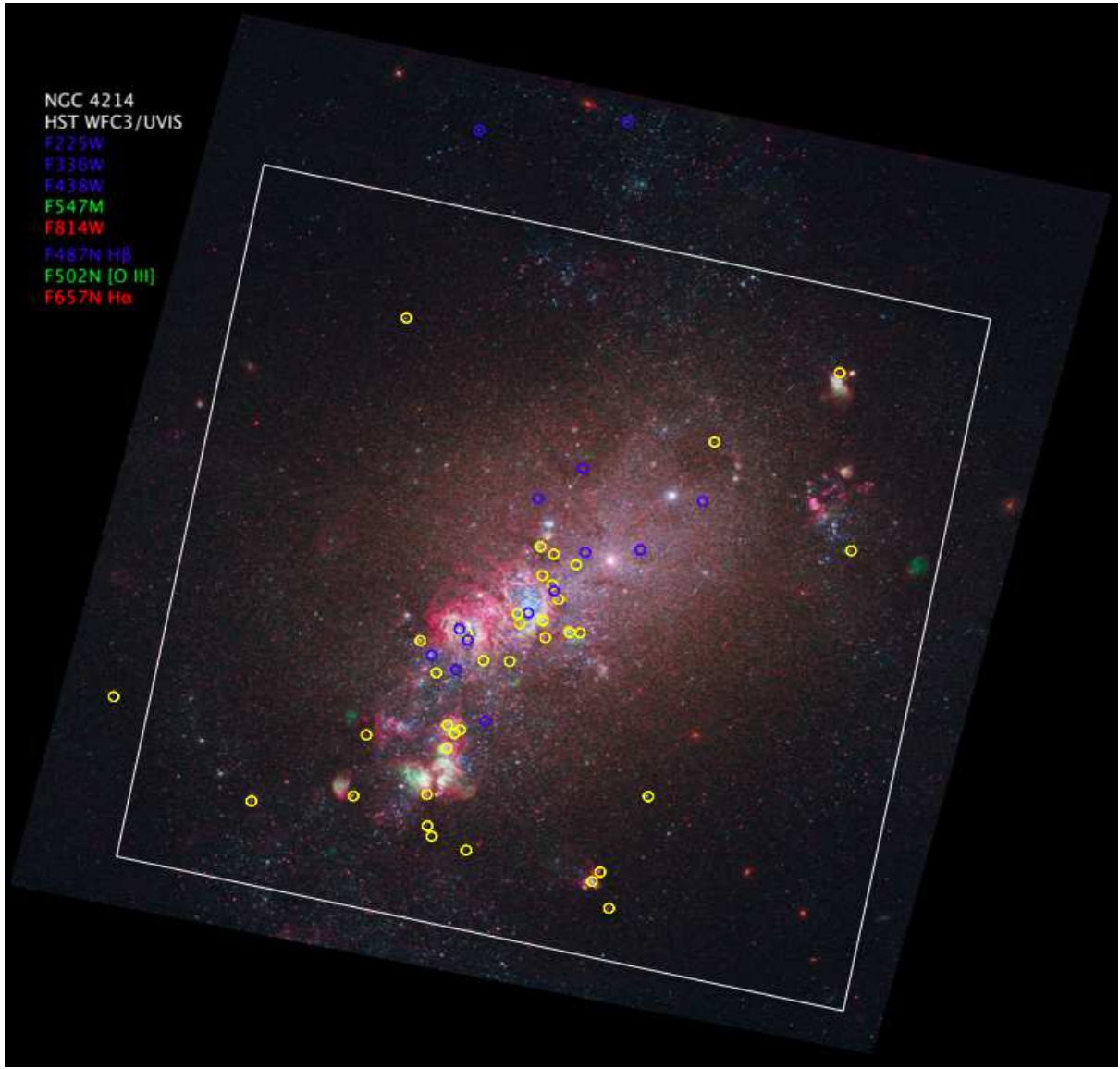} 
   \caption{Color composite WFC3 image of NGC 4214 courtesy of Zoltan Levay (STScI-2011-14), R. O'Connell (GO 11360), and the WFC3 SOC. The footprint of the WFC3/IR images is shown in white.  Blue circles are clusters with undetected H$\alpha$ emission, yellow circles are clusters with H$\alpha$ emission. There are a total of 52 compact clusters.}
   \label{fig:clusterpos}
\end{figure*}

As demonstrated in \citet{2010ApJ...719L.158C}, it is possible to constrain the upper end of the IMF in external galaxies without the use of individual star counts. Instead, a nearly coeval population that still contains the most massive members capable of producing ionizing photons can be constructed from the sum of unresolved young clusters in the galaxy. Therefore, measuring Q(H$^{0}$), the hydrogen ionizing photon rate, in these young clusters is equivalent to measuring the number of massive stars. This method relies on normalizing the ionizing photon rate to the age-independent cluster mass, which is an extension of the method described in \citet{2009A&A...495..479C}.   The treatment presented here can eliminate the need for age-dependent bolometric luminosities, but does require that the cluster ages are accurately determined.  The pilot study done on M51a by \citet{2010ApJ...719L.158C}, found that there was no obvious dependence of the upper mass end of the IMF on the mass of the star cluster down to $\sim$ 10$^{3}$ M$_{\sun}$, but a more extensive analysis including additional populations than M51a is needed for conclusive evidence. Specifically galaxies with star formation rates (SFR) below the threshold for which IMF variances have been suggested ($\leq$ 0.1 M$_{\sun}$ yr$^{-1}$) need to be investigated. This paper aims to extend this study using a nearby galaxy with a lower SFR more similar to those dwarf galaxies which may be exhibiting a deficiency of ionizing photons.

NGC4214 is an irregular, LMC-type, star bursting galaxy located $\sim$3 Mpc away \citep{2010Ap&SS.330..123D} with an H$\alpha$ and UV SFR of 0.16 M$_{\sun}$ yr$^{-1}$ and  0.22 M$_{\sun}$ yr$^{-1}$  respectively \citep{2011ApJS..192....6L,2008ApJS..178..247K,2009ApJ...706..599L} as well as a sub-solar (Z$\sim$ 0.25 Z$_{\sun}$) metallicity \citep{1996ApJ...471..211K}. The star formation history (SFH) of the central region of NGC4214 shows a strong increase in SFR starting 100 Myr ago with a prominent peak at recent times ($\leq$ 10 Myr); despite this, less than 1\% of the mass of the galaxy is due to the current star formation event \citep{2011ApJ...735...22W}.  Due to its proximity and recent star forming activity, NGC 4214 is an ideal test-bed for the upper end of the IMF (uIMF). In Section 2 of this paper we will discuss the observations and cluster selection criteria, in Section 3 we will present the models and age and mass determinations, and in Section 4 we discuss the results.

\section{Observations and Cluster Selection}
Observations were taken with HST WFC3/UVIS and WFC3/IR as part of GO 11360 (PI: O'Connell). The observations on which we concentrate here include F225W (1665s), F336W (1683s), F438W (1530s), F547M (1820s), F657N (1592s),  F814W (1339s), F 110W (1198s), and F128N (1198s), shown in Figure \ref{fig:clusterpos}. For ease of discussion, we will refer to these as \textit{NUV}, \textit{U}, \textit{B}, \textit{V}, \textit{H$\alpha$}, \textit{I}, \textit{J}, and \textit{P$\beta$} respectively. Each flat-fielded image was co-added, cosmic rays were removed, and corrections for distortion were made using the task MULTIDRIZZLE into a final pixel scale of 0$\arcsec$.0396 pixel$^{-1}$. This corresponds to 0.58 pc  pixel$^{-1}$ at a distance of 3 Mpc. See \cite{2010Ap&SS.330..123D} for a full explanation of the reduction procedure, including the creation of continuum subtracted H$\alpha$ images.  A similar procedure was adopted to create a continuum subtracted P$\beta$ image, which covers $\sim$75$\%$ of the  UVIS field of view.  The footprint of the IR field of view is also shown as a white outline in Figure \ref{fig:clusterpos}.

Cluster candidates were identified using a technique similar to that described in \citet{2011ApJ...727...88C,2010ApJ...719..966C} for M51 and M83. Aperture photometry was performed on the wide and medium band images using the IRAF task PHOT with an aperture of 3 pixels, and a background annulus between 10 to 13 pixels.  The aperture corrections were addressed similarly as was done on clusters in M83 from \citet{2010ApJ...719..966C}.  In their study they chose two different methods for aperture correction, both of which rely on the concentration index ($C$, the difference in magnitudes between 3 pixel and 0.5 pixel radius).  Method one used a single value for the aperture corrections of point sources ($C$ $ <$ 2.3) and a slightly larger value for extended sources ($C$ $>$ 2.3).  For this sample we chose to use their second approach, which is to use an aperture correction equation for extended objects with 2.3 $< $C$ \leq$ 3.4. Photometric conversion from counts s$^{-1}$ to erg cm$^{-2}$ s$^{-1}$ were accomplished using the filter dependent PHOTFLAM values provided by the STScI website.  Galactic foreground extinction of $E(B-V)$ = 0.02 was corrected using the Milky Way extinction curve from \citet{1999PASP..111...63F}.
 
 Due to the more extended nature of HII regions surrounding the stellar clusters, aperture sizes that scaled with the cluster mass according to the expected Str\"{o}mgren radius were used to measure the hydrogen recombination lines on the continuum-subtracted $H\alpha$+[N II] and $P\beta$ images. As was done in \cite{2010ApJ...719L.158C}, a radius of about 0.35 R$_{Stromgren}$ was selected due to the crowding of the clusters, which corresponds to a range between 6-10 pixels.  This radius is sufficiently larger than the PSFs for both the H$\alpha$ and P$\beta$ images, so there are no concerns of PSF variations. The local background was subtracted using a 3 pixel wide annulus centered on the cluster outside of the aperture radius in order to avoid contamination from other diffuse emission.  Aperture corrections were calculated from a few, very isolated sources, and were applied to the other regions. We found that an additional correction of 1.30 times the flux was needed to go from a 20 pixel aperture to an ``infinite'' aperture.  Contamination from [N II] was removed using the average galactic [N II]/H$\alpha$ ratio of 0.11 from \citet{1996ApJ...471..211K}.  The corrections for the ionized gas extinctions were measured region by region using the WFC3/IR P$\beta$ image for those regions where the IR and UVIS overlap (see Figure \ref{fig:clusterpos}) using the formulation in \citet{2000ApJ...533..682C} and were applied to the H$\alpha$ luminosities. 
 
The cluster catalog was populated by the combination of two methods.  The first, and most robust, was using the automated method discussed in depth in \citet{2010ApJ...719..966C} which accounts for roughly 80$\%$ of the cluster catalog. To make sure all clusters are accounted for, we also use a manual selection procedure from a careful examination of the WFC3 images.  This ensures we identify clusters in crowded regions or clusters near a bright star which may have been missed in the automated process.  In total we have identified $\sim$ 400 cluster candidates.

\section{Analysis}
In this section we will present an in-depth discussion of the analysis procedure used in this paper, and a detailed description of the filter convolution is included in the Appendix.  For the mass and age determination of the clusters we have used broad-band photometry (without the H$\alpha$ included) to determine ages and masses of the clusters using both a canonical and truncated IMF from stochastic and deterministic stellar models.  The best fit model-derived masses of clusters with ages less than 8 Myr are then binned into three distinct mass bins.  Within these mass bins the masses are summed and the H$\alpha$ luminosity are summed to determine the  L$_{H\alpha}$/M$_{cl}$ ratios for various cluster masses ($<\frac{L_{H\alpha}}{M_{cl}}>  =\frac{\sum_{i}L_{H\alpha i}}{\sum_{i}M_{cli}}$).  These ratios are then compared to predicted models with two different assumptions about the IMF, the canonical one and the variable upper mass limit \citep[see Discussion]{2003ApJ...598.1076K}, in order to constrain the IMF of NGC 4214.

\begin{figure}[h] 
   \centering
   \includegraphics[width=3.3in]{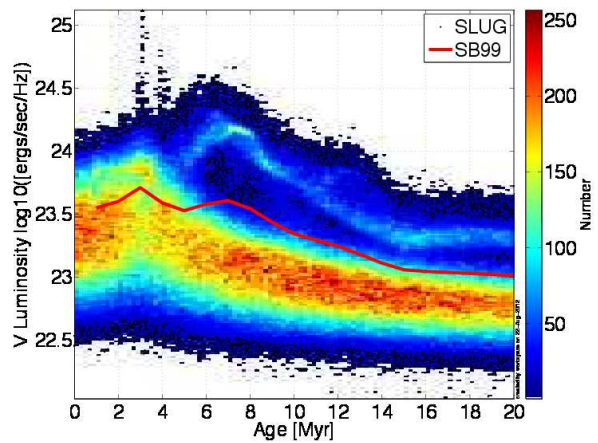} 
   \caption{Comparison of V-band luminosities of 1 $\times$ 10$^{3}$ M$_{\sun}$ SLUG models from 1-20 Myr with SB99 (red line) with A$_{V}$ = 0.  The magnitude of scatter from the stochastic models will also create a scatter of cluster mass. Note that the arithmetic mean of the SLUG models match the SB99 models, but that the geometric mean is slightly lower.}
   \label{fig:lumscatter}
\end{figure}

\subsection{The Models}
In order to accurately determine ages and masses of the clusters, the spectral energy distributions (SEDs) created from the photometry of the $NUV,U,B,V, I$ observations were compared to stellar synthesis models. While the norm has been to use deterministic models such as STARBURST99 (SB99) \citep{1999ApJS..123....3L}, these models assume a fully sampled stellar IMF, which for smaller mass clusters could lead to the inclusion of unphysical fractions of stars.  With stellar clusters of high masses ($\geq$ 1 $\times$ 10$^{4}$ M$_{\sun}$), we expect the IMF to be fully populated \citep{2006ApJ...648..572E,2012ApJ...745..145D}, so this is not a problem, but, as the cluster masses decrease, massive stars are less likely to be formed and massive stellar populations are not fully represented. In order to properly measure parameters for low mass clusters ($\sim$500 - 5000 M$_{\sun}$), it is then important to turn to stochastic modeling. For this paper we will mainly focus on the stochastic models of SLUG \citep[Stochastically Lighting Up Galaxies]{2012ApJ...745..145D} which performs the synthesis of composite populations using individual stellar clusters which are stochastically populated with stars, using the IMF as a probability distribution function.  Other stochastic models are presented in \citet{2010ApJ...724..296P}, but are not used in this study. According to \citet{2006A&A...451..475C,2004A&A...413..145C} clusters with masses below 10$^{3}$ M$_{\sun}$ and ages less than 10$^{7}$ years may be susceptible to color biases from deterministic stellar synthesis models.  NGC 4214 has numerous small clusters making it pertinent that stochastic models be used.  As a check and comparison, we also employ the deterministic SB99 models using the same input parameters and use outputs which contain both stellar and nebular emission as cautioned by \citet{2010ApJ...708...26R}.

Both sets of models use a Kroupa IMF between 0.08-120 M$_{\sun}$ \citep{2001MNRAS.322..231K}, Padova AGB tracks with z=0.004, and assume that the clusters form in a single instantaneous burst.  For the truncated SLUG models, where the maximum mass was only allowed to be 30 M$_{\sun}$, a Salpeter IMF was used. The SB99 models contain ages between 1-200 Myr in time steps of every 1 Myr for models between 1-20 Myr, and every 25 Myr for models between 25-200 Myr (to easily distinguish old clusters.) New SB99 models were generated between 2-8 Myr with time steps of 0.2 Myr for more accurate comparison with SLUG models. The SLUG models include about 40000 cluster templates with ages from .01 Myr to 20 Myr with a cluster mass of 1 $\times$ 10$^{3}$ M$_{\sun}$ (Figure \ref{fig:lumscatter}).  Tests have been run on a subset of clusters using 5 $\times$ 10$^{2}$, 1 $\times$ 10$^{3}$, and 3 $\times$ 10$^{3}$ M$_{\sun}$ SLUG models and we have found that the differences in inferred ages and masses of the same cluster among the various models is small, and is already encompassed by the uncertainties generated within one 1 $\times$ 10$^{3}$ M$_{\sun}$ model, which is described below.  Therefore, to simplify the analysis and most accurately reflect the masses of the clusters in NGC4214, we use only 1 $\times$ 10$^{3}$ M$_{\sun}$ SLUG models for our cluster sample. SLUG models do not allow for binarity, but a recent study by \citet{2012MNRAS.422..794E} indicates that at masses $\geq$ 10$^{3}$ M$_{\sun}$, the scatter in L$_{H\alpha}$/M$_{cl}$ is the same between models that use single stars and those that introduce binaries, so this should not introduce additional uncertainties.

\subsection{Age and Mass Determination}
To estimate the age, mass, and extinction of each cluster we employ a reduced $\chi ^{2}$ fitting technique between both the SLUG and SB99 models and the cluster photometry. To do this, we have used the Yafit (Yet Another Fitting Tool \footnotemark{\footnotetext[10]{http://www.star.bris.ac.uk/~mbt/yafit/}}) curve fitting tool, which was created to fit photometry with model SEDs.  It provides both a graphical and a numerical output containing the reduced $\chi ^{2}$ value and the scaling factor between model and observation. The ensemble of photometry for each cluster was compared to both SLUG and SB99 models spanning the complete reddening range between 0$\leq$ $E(B-V)$ $\leq$0.40.  Note, we do not use the H$\alpha$ filter as part of the fit, as we do not want to bias our sample based on the presence or absence of H$\alpha$ emission.

\begin{figure*}[t!]
\centering
\includegraphics[width=3.1in]{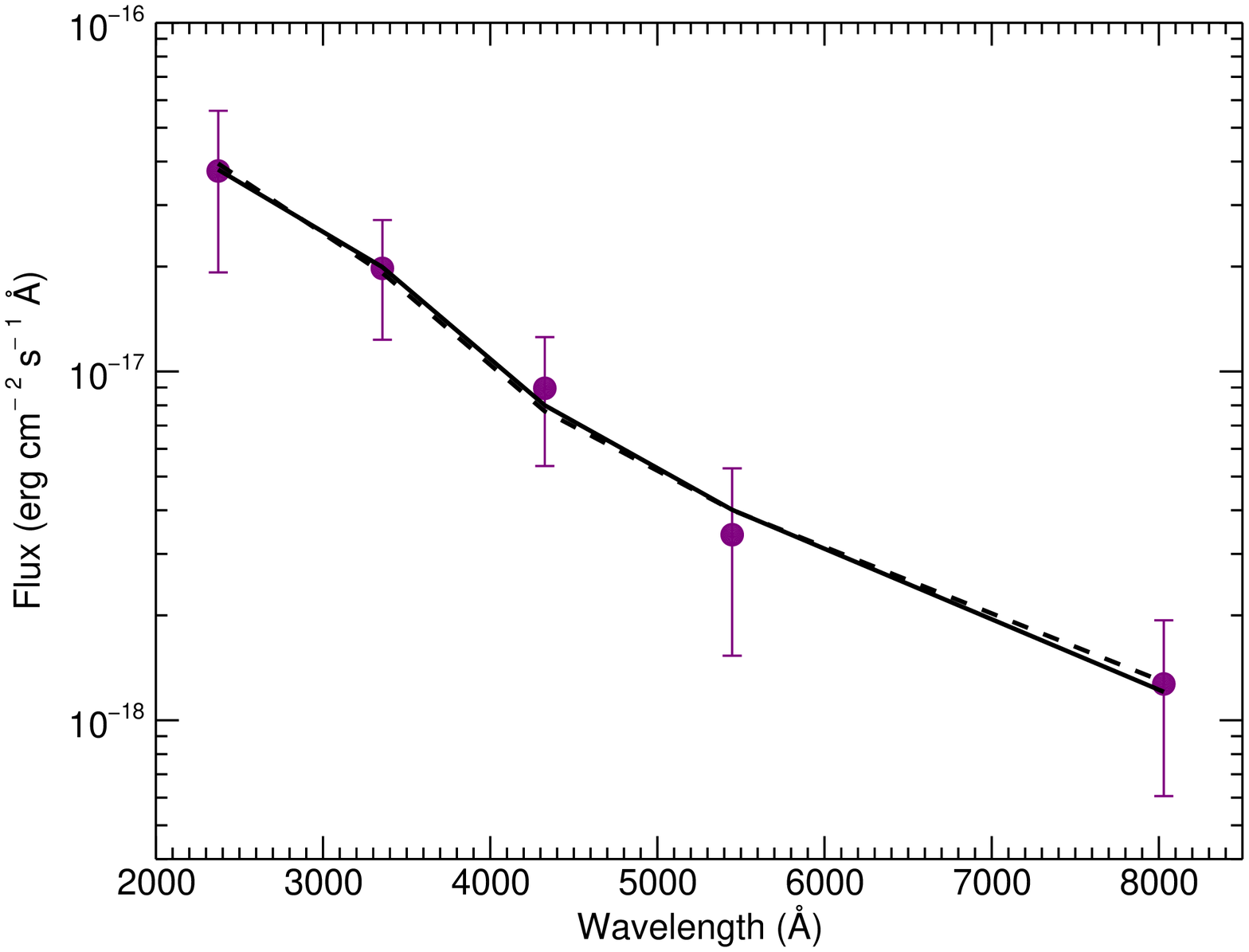}
\
\includegraphics[width=3.1in]{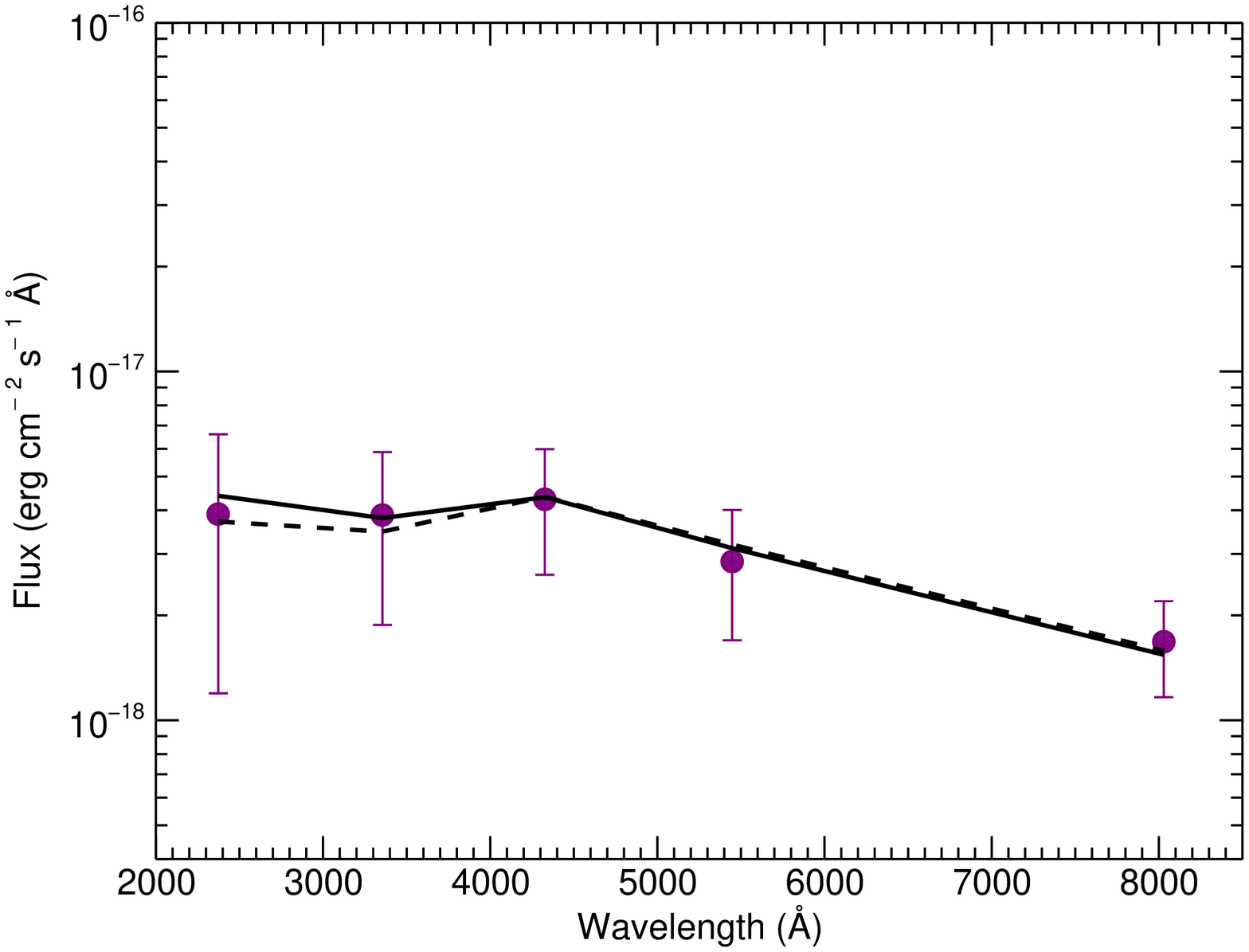}
\
\includegraphics[width=3.1in]{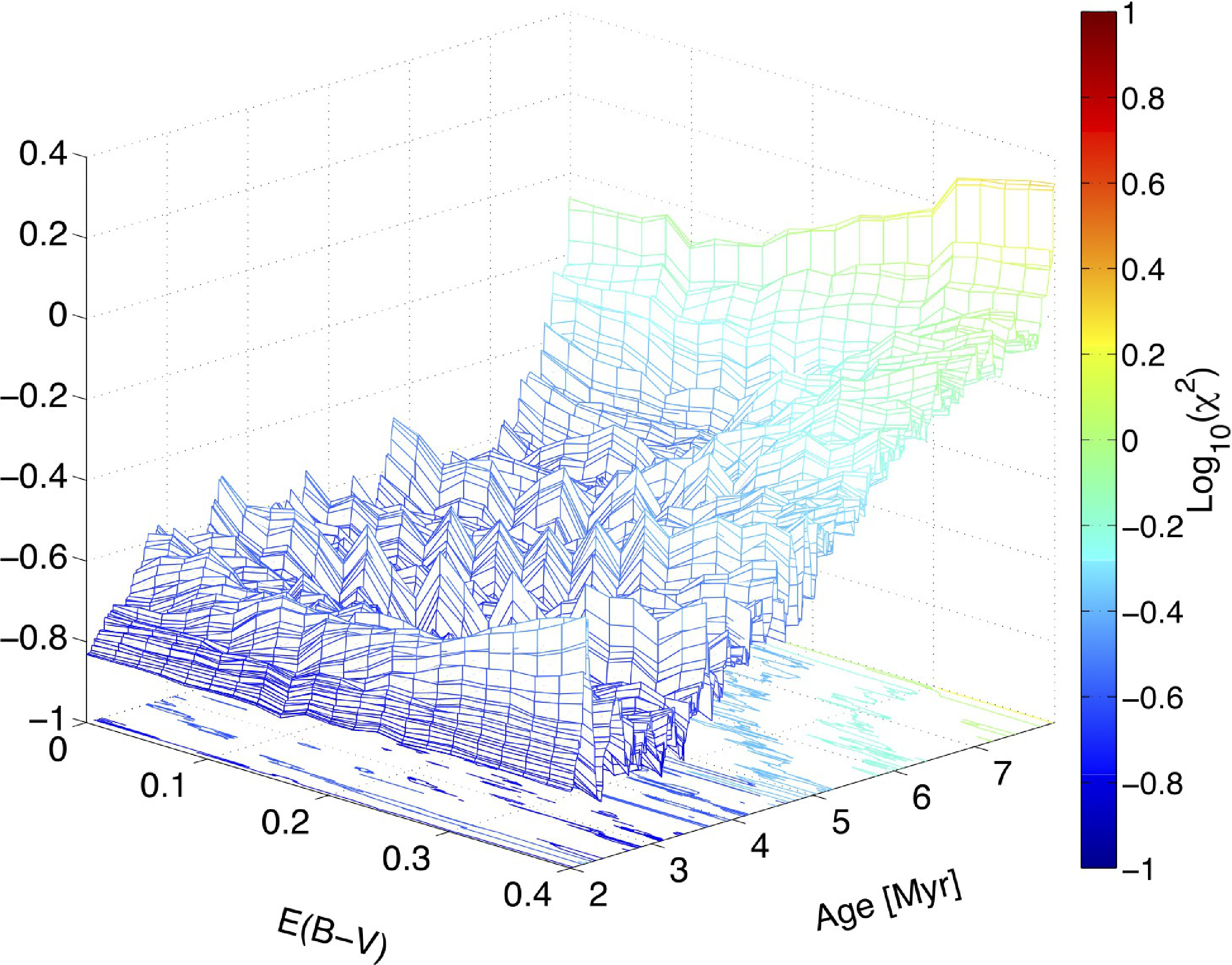}
\
\includegraphics[width=3.1in]{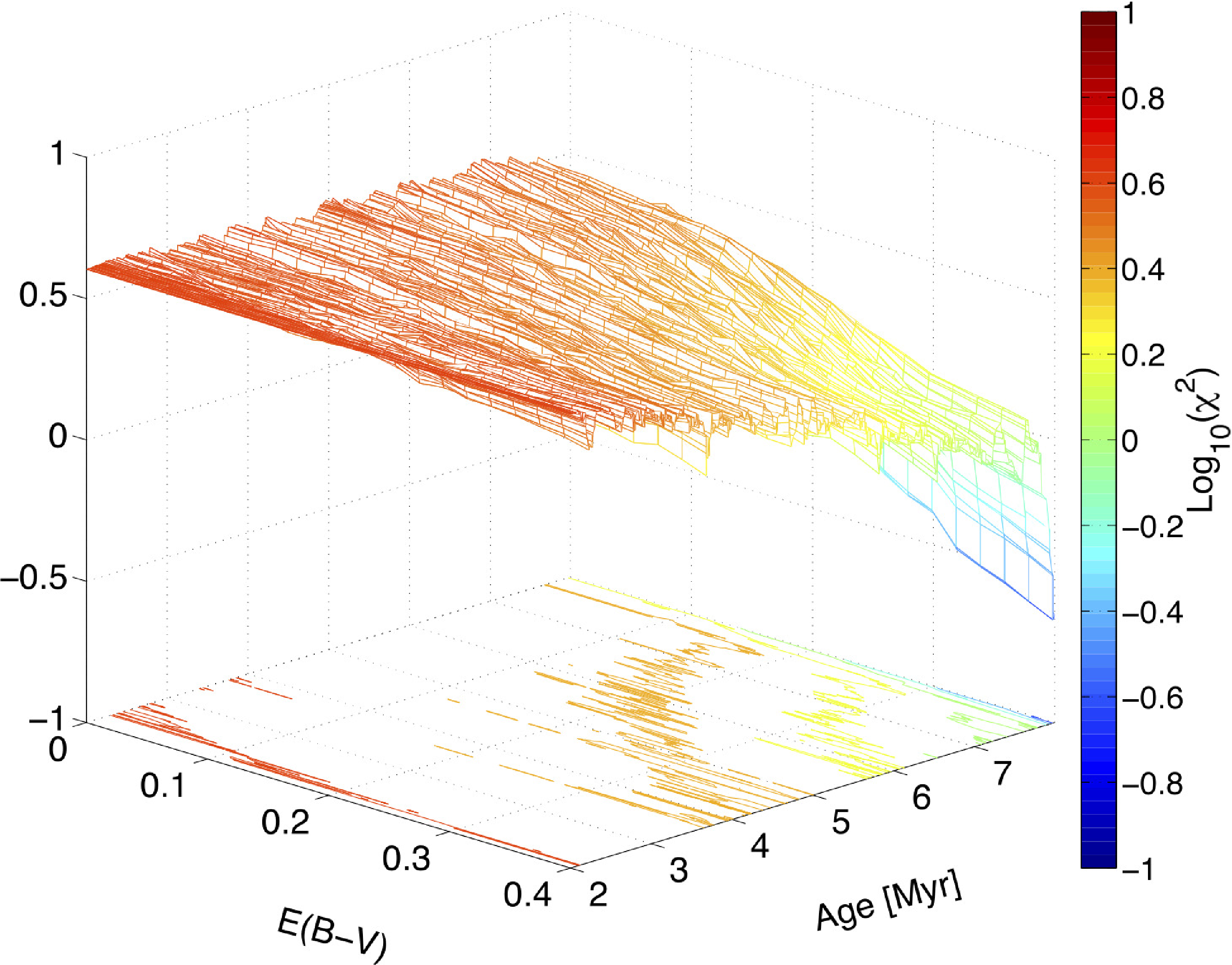}
\
\caption{ \bf{Top: SED model fits for a $\sim$2 Myr cluster (left, cluster 402 in Table 1) and a $\sim$7.5 Myr cluster (right, cluster 188 in Table 1).  Each fit uses an $E(B-V)$ of 0.06, and 0.40, respectively. SLUG models are indicated by a solid line, and SB99 models by a dashed line. The large uncertainties in the photometry can allow for a large range of fits with an acceptable $\chi ^{2}$ value.  Bottom: Contours for the reduced $\chi ^{2}$ values for various age and attenuation values for a $\sim$2 Myr cluster (left) and a $\sim$7.5 Myr cluster (right).   }}
\label{fig:fits}
\end{figure*}

After all of the observations were compared against each model, the fits for each cluster were then organized into increasing reduced $\chi ^{2}$ values. It is important here to point out that there is no single solution for the age and extinction of the cluster, but instead there is a range of best fits which could produce the model fit. In Figure \ref{fig:fits} we show both the SED fits and the range of ages and extinctions spanned by the model fits to the youngest ($\sim$ 2 Myr) and oldest ($\sim$ 7.5 Myr) cluster in our sample. By using the large range of ages and extinction consistent with the model fits, the actual mass distribution has been spread over a large range of values and only produces a peak at the most probable value. Therefore, we allowed the $\chi ^{2}$  values to range between 0 and 1, as was done in \citet{2003MNRAS.345..161P}, and include all ages, extinctions, and therefore corresponding masses within that range. Masses were determined using the scaling factor output by Yafit and attributing error bars consistent with the error bars in the ages and extinction values.  As with \citet{2010ApJ...719L.158C}, we also find that changing the upper mass limit of the IMF in both the SLUG and SB99 models from 120 to 30 M$_{\sun}$ increases the mass estimates by roughly 2.5. To double check our initial results, comparisons with previous studies which have determined ages and masses of some of the clusters in NGC 4214 were conducted and found to be comparable.  For example, the young massive cluster located at the center of NGC4214 (R.A.=12$^{h}$15$^{m}$39$^{s}$.44, Dec = 36$^{\circ}$19$^{\prime}$34$^{\prime\prime}$.94), noted here as Cluster 1, has age and mass estimates of 4-5 Myr and 2.7$\pm$0.4 $\times$ 10$^{4}$ M$_{\sun}$ \citep{2000AJ....120.3007M,1996ApJ...465..717L} using a Salpeter IMF between 1-100 M$_{\sun}$.  Our best estimate, from the best fit from both SLUG and SB99, is 4.2 and 4.8 Myr, respectively, with a corresponding best fit mass of 9.3$\pm$4  and 9.7$\pm$3 $\times$ 10$^{4}$ M$_{\sun}$, which is consistent with these previous studies within 2$\sigma$. Unfortunately the cluster has blown an extensive asymmetric bubble, clearing the surrounding region of much of its hydrogen gas and allowing more than half of the ionizing photons to escape \citep{2000AJ....120.3007M}, making accurate measurements of L$_{H\alpha}$ extremely difficult.  For completeness purposes, we will use the H$\alpha$ flux for Cluster 1 (I-As) quoted in \citet{2000AJ....120.3007M}, which gives an H$\alpha$ luminosity of 8.4 $\times$ 10$^{37}$ erg s$^{-1}$.

\begin{figure*}[t!] 
   \centering
   \includegraphics[width=2.5in, angle=270]{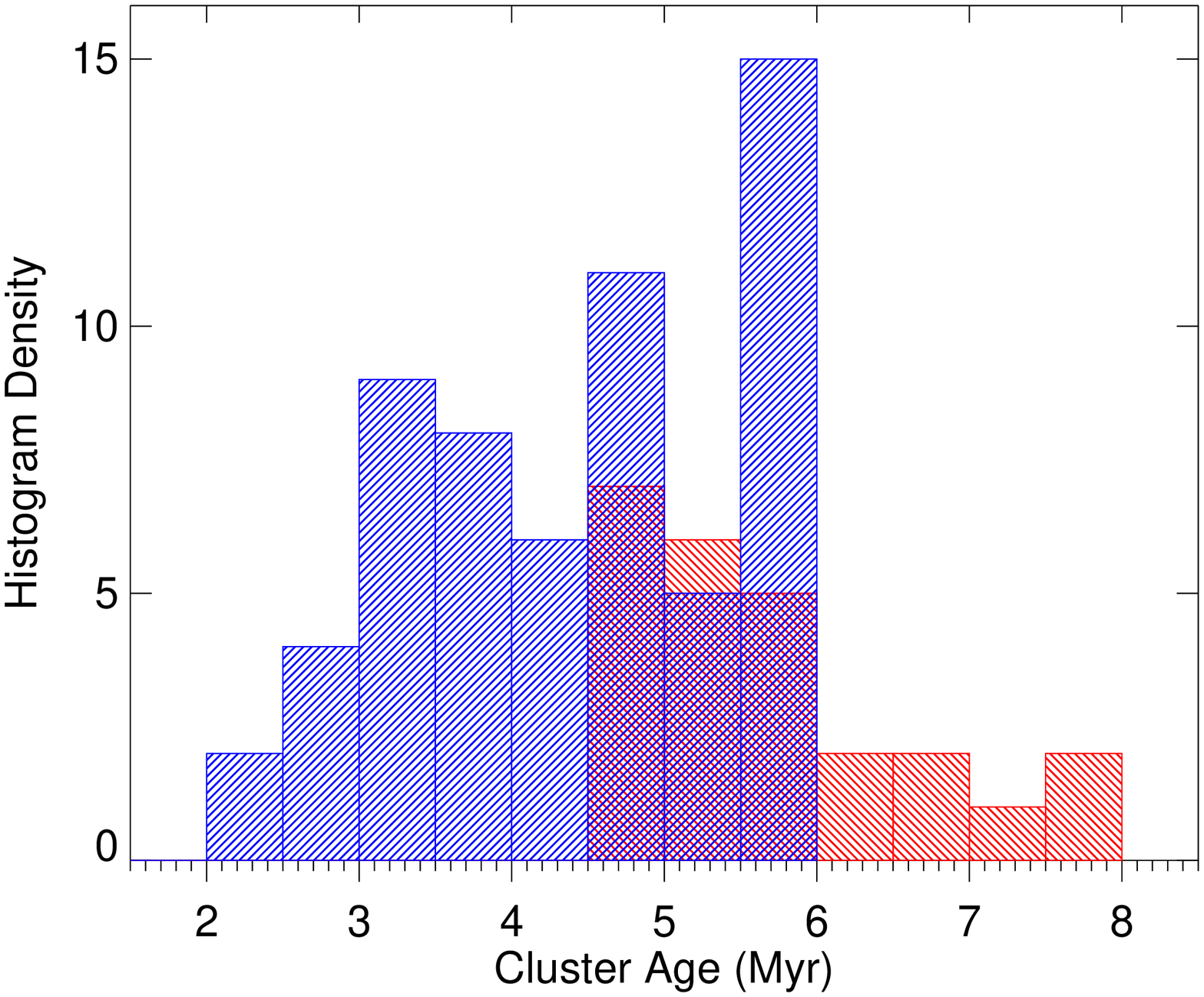} 
   \
   \includegraphics[width=2.5in, angle=270]{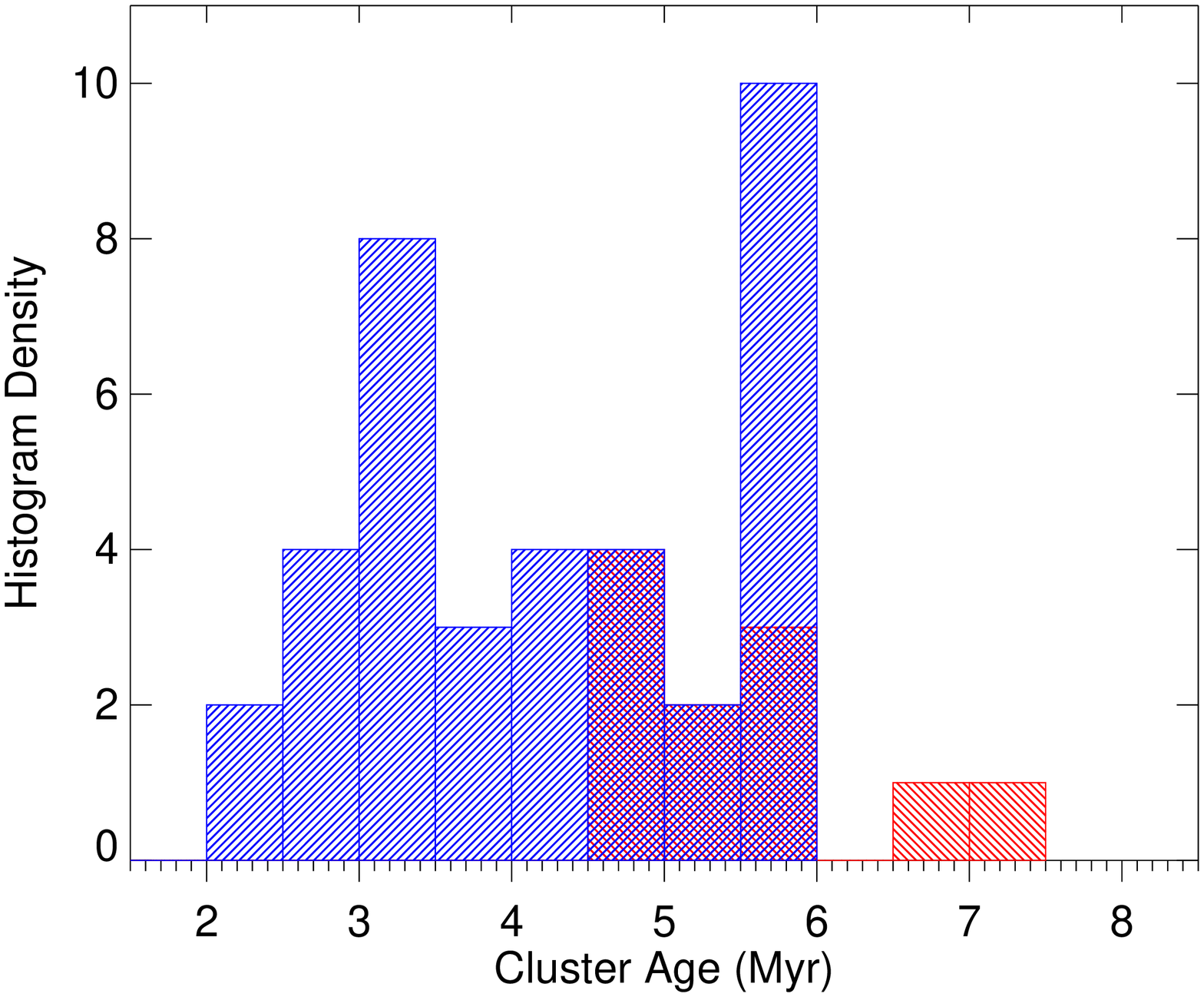}
   \caption{Histograms of relative frequencies of best fit cluster ages for all young clusters (left) and clusters with masses $>$ 500 M$_{\sun}$ (right).  The blue histogram indicates those clusters with measured H$\alpha$ emission, while the red indicates those with only upper limits.  From both plots it is apparent that clusters $<$ 4.5 Myr will still retain H$\alpha$ emission, while those $>$ 6 Myr will not. Cluster with ages between 4.5-6 Myr have a lower probability of producing H$\alpha$ emission. These plots exclude those clusters with masses $>$ 10$^{4}$ M$_{\sun}$ as they may not be fully explored by the SLUG models and are far more likely to have expelled their surrounding hydrogen gas at a much younger age.}
   \label{fig:agewithha}
\end{figure*}

By $\leq$ 8 Myr, the compact HII region surrounding stellar clusters has expanded into a shell structure which disperses into the ISM \citep{2011ApJ...729...78W} and massive stars capable of producing ionizing photons ($>$ 15-20 M$_{\sun}$) have disappeared. For example, SB99 models using the parameters listed above, show that the H$\alpha$ luminosity at 8 Myr is only 2.5$\%$ of L$_{H\alpha}$ at 2 Myr. Therefore it is necessary to exclude all clusters $>$ 8 Myr in order to keep objects in which the HII region is still density bound and the massive stars are still retained.  This ensures that uncertainties in the ionizing photon rate are reduced and that constraints can still be made on the upper end of the IMF. As will be discussed below and is shown in Figure \ref{fig:agewithha}, including ages greater than 6 Myr may already be too old for this type of analysis. We must also be careful of confusion, for example, including multiple objects which may share the same HII region. For this reason, we have excluded those clusters which may share in ionizing common gas, removing clusters which are in excessively crowded regions which would hinder the correct measurement of  L$_{H\alpha}$. This has left us with a total of 89 clusters between the ages of 2-8 Myr, 52 of which have masses that are $\geq$ 500 M$_{\sun}$, and either have a measured H$\alpha$ luminosity (38) or have an H$\alpha$ luminosity that is non-detectable down to the 3$\sigma$ limit of 1.6 $\times$ 10$^{35}$ erg s$^{-1}$. Ten of these objects (Table 1, footnote $c$) have PSFs that are consistent with a single massive star, yet their SEDs require the presence of multiple stars to be fully accounted for in flux.  These low-multiplicity clusters tend to be among our lowest mass systems (Table 1), and may be the NGC 4214 equivalent of the Trapezium cluster in Orion, which has a 1-1.5 pc size and a handful of 15-30 M$_{\sun}$ stars. Therefore, they have still been included in the sample and are indicated in Table 1 with the rest of the clusters.  Table 1 also includes the corresponding ages, extinctions, and masses from both SLUG and SB99 for all 52 clusters; Figure  \ref{fig:clusterpos} shows their placement in the galaxy. We have also included the cluster naming nomenclature used in \citet{2000AJ....120.3007M} where applicable for ease of comparison. 

\begin{figure}[h] 
   \centering
   \includegraphics[width=3.3in]{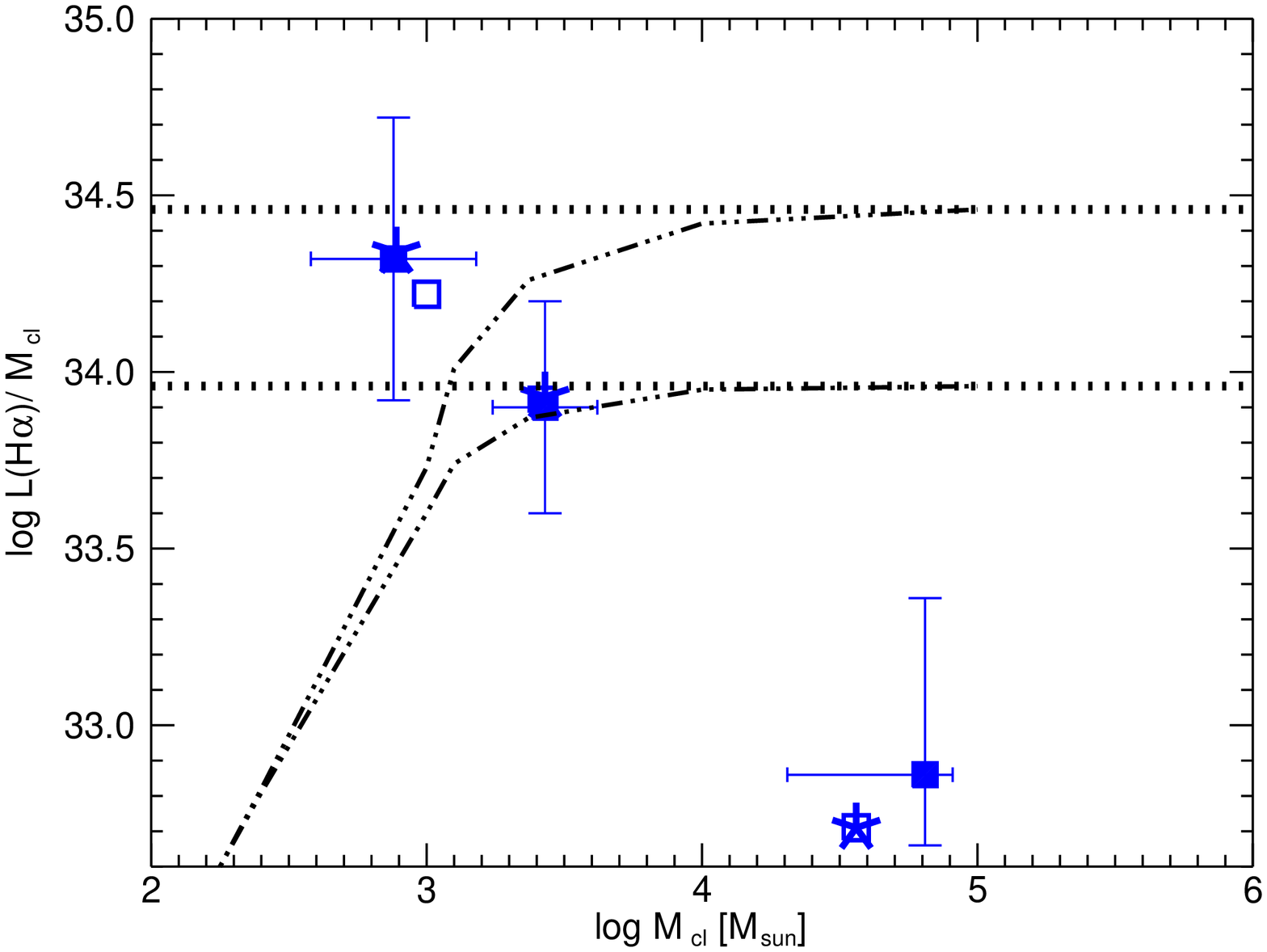} 
   \includegraphics[width=3.3in]{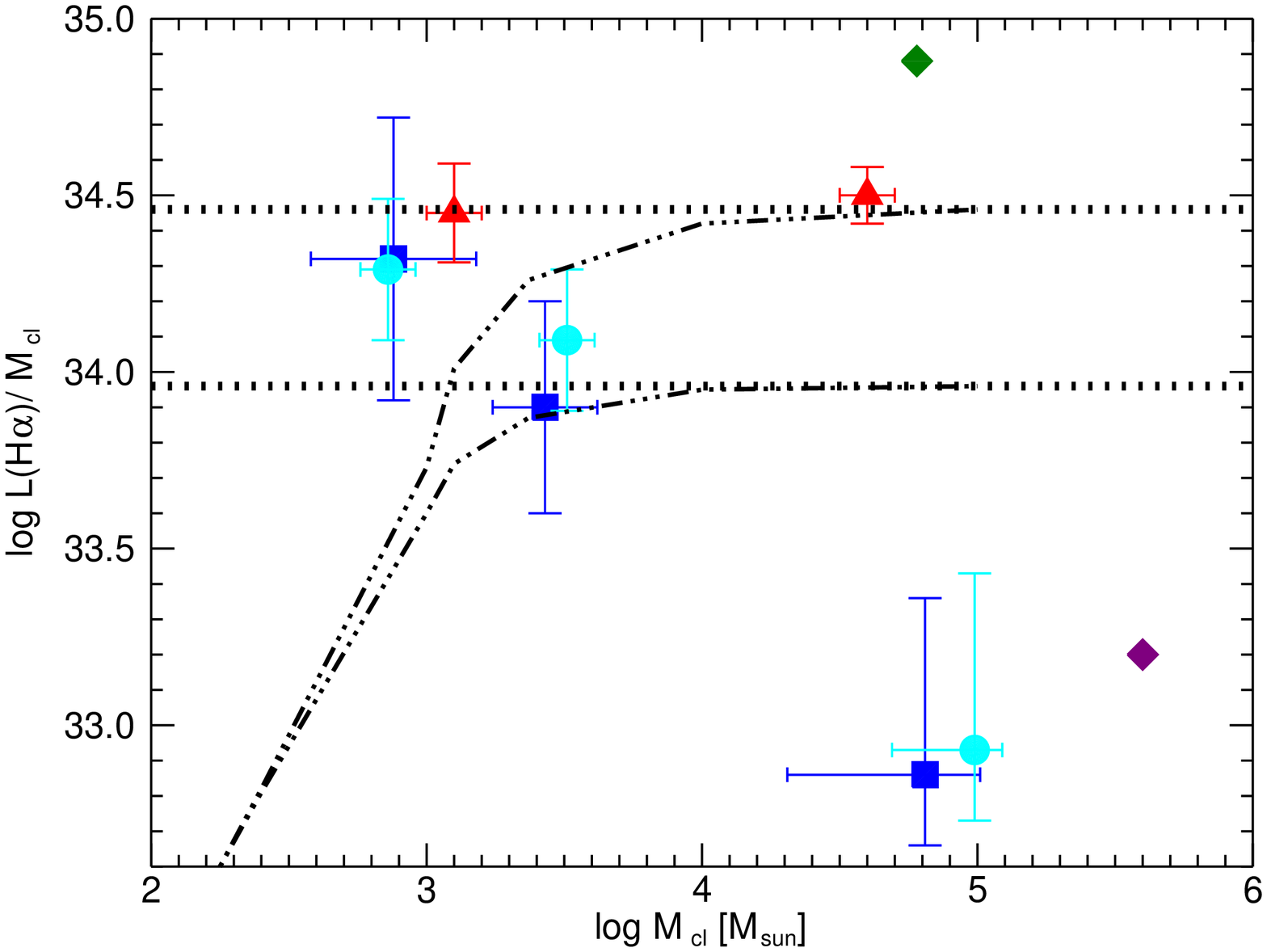} 
   \caption{ Top: Location of L$_{H\alpha}$ /M$_{cl}$ for mass bins in NGC 4214. Filled blue squares only include clusters with H$\alpha$ emission (all of which are less than 6 Myr), while open blue squares include all clusters, and blue stars include all clusters without those objects which could possibly be single stars or low-multiplicty clusters. Dotted line is the expected L$_{H\alpha}$ /M$_{cl}$ for a universal IMF while the dash-dotted line is for a variable upper mass limit \citep{2003ApJ...598.1076K,2010MNRAS.401..275W}  where the most massive star in a cluster is determined by cluster mass. The top lines show averaged models between 2-5 Myr, while the bottom is between 2-8 Myr.  Bottom:  Same as top, but also containing cluster measurements of NGC 4214 from SB99 (cyan circles), from M51  \citep[red triangles]{2010ApJ...719L.158C}, R136a (\citet{2012ApJ...755...40P}, green diamond) and NGC 330 (\citet{2012ApJ...755...40P}, purple diamond). The location of the largest mass blue and cyan symbols is highly impacted by the uncertainty of the H$\alpha$ luminosity of Cluster 1.}
   \label{fig:binnedplot}
\end{figure}

\section{Discussion}

If the IMF is populated purely stochastically, then we can expect that 100 10$^{3}$ M$_{\sun}$ clusters would contain the same numbers and masses of stars as one 10$^{5}$ M$_{\sun}$ cluster, and that both would represent a fully sampled IMF \citep{2001ASPC..243..255E,2006ApJ...648..572E}.  On the other hand, if the most massive star in a cluster is limited by the mass of the parent cluster we might not expect any high mass stars to be present in 1000 M$_{\sun}$ clusters. For example, in the variable upper mass limit formulation of \citet{2010MNRAS.401..275W}, they propose a M(max)$_{*}$ -- M$_{cl}$ relation in which no stars more massive than 35 M$_{\sun}$ would be present in a 10$^{3}$ M$_{\sun}$ cluster. The summation of the total ionizing flux from these small clusters divided by the total cluster mass should then be much lower than the ionizing flux from the single large cluster divided by its mass, and as cluster mass decreases there is a deviation from the ratio of ionizing photons to mass expected by a universal IMF (Figure \ref{fig:binnedplot}, dashed-dotted line). In a universal IMF scenario, then, the summation of the total ionizing flux from the small clusters divided by the mass will be consistent with that of a single large cluster. Even though most low mass ($\leq$ 500 M$_{\sun}$) clusters will produce low H$\alpha$ luminosities in the universal scenario, there will be some that do produce a large ionizing continuum from the odd star well over 20 M$_{\sun}$. \cite{2010A&A...522A..49V} estimates that only 20$\%$ of 100 M$_{\sun}$ clusters will have stars large enough to create an H II region, therefore given a large enough sample the effects will average out.  By summing the L$_{H\alpha}$ and masses of all of the small clusters into one data point, not only are the observational uncertainties reduced but the stochastic effects are minimized.

We have used three mass bins (see Figure \ref{fig:binnedplot}), each with a mean mass of 7.5  $\times$ 10$^{2}$ M$_{\sun}$, 2.2 $\times$ 10$^{3}$ M$_{\sun}$, and 4 $\times$ 10$^{4}$ M$_{\sun}$.  The error bars in each bin are obtained by adding in quadrature the individual mass and luminosity errors of each cluster fit. The largest mass bin has only 5 members in total, including the super star cluster at the center of the galaxy (Cluster 1), for which the H$\alpha$ luminosity is highly uncertain and we are using a value obtained in \citet{2000AJ....120.3007M}. These extremely large clusters with negligible H$\alpha$ emission are included to illustrate the possibility that we may have feedback occurring in our larger clusters which is dispersing the gas more efficiently. Indeed we note that the most massive star clusters in NGC 4214 are surrounded by ionized gas shells (see below). We have also included for illustration those clusters that have masses greater than 500 M$_{\sun}$, but only upper limit measurements for L$_{H\alpha}$ (1.6 $\times$ 10$^{35}$ erg s$^{-1}$, open squares).  Finally, we have included the full sample of objects with masses greater than 500 M$_{\sun}$ with those objects which may be single stars or low metallically clusters removed. In the lower mass bins, the error is dominated by the range of masses. Also plotted in Figure \ref{fig:binnedplot} is the expected average L$_{H\alpha}$/M$_{cl}$ from a 1/5 Z$_{\sun}$ SB99 metallicity model that is fully populated up to 120 M$_{\sun}$ between 2-5 Myr (top dashed line), and 2-8 Myr (bottom). The expected range for a M$_{*}$-- M$_{cl}$ model where the most massive star in the cluster is a function of cluster mass \citep{2010MNRAS.401..275W} is shown in the dashed-dotted line also averaged between 2-5 Myr (top) and 2-8 Myr (bottom).   If the metallicity of NGC 4214 is 1/4 Z$_{\sun}$ we do expect the data to fall somewhat below the stellar synthesis models, since lower metallicities create higher H$\alpha$ luminosities.

When including upper limits in H$\alpha$ the clusters in the highest mass bin have a value of L$_{H\alpha}$/M$_{cl}$ considerably lower than that of lower mass clusters (Figure \ref{fig:binnedplot}). We have investigated whether the clusters in the highest mass bin may be experiencing feedback effects which have caused expulsion of the gas from the HII region.  \citet{2012ApJ...755...40P} have found that the amount of ionizing photons lost from a cluster can be dependent on the HI density surrounding the cluster. With this in mind, we have used the HI maps of NGC 4214 published in \cite{2001AJ....121..727W} to locate the clusters in the largest mass bin without H$\alpha$ emission.  We find that they are located in regions that have HI densities roughly 2/3 that of the maximum density, which by itself would not indicate that the clusters should experience significant ionizing photon loss.  When observing the clusters in the H$\alpha$ image though, it is quite clear that some mechanism has blown much of the gas away from the clusters and they are surrounded by wind blown bubbles. This is not completely unexpected, as these clusters with extremely low L$_{H\alpha}$ values are located in the NGC 4214-I region, which \citet{1998A&A...329..409M} found there was a significant decoupling of the stellar clusters with the ionized gas.

In Figure \ref{fig:agewithha} we present histograms of the ages of the stellar clusters with detected H$\alpha$ (blue) and non-detected H$\alpha$ (red) in our sample, including one containing all clusters younger than 8 Myr with masses $<$ 10$^{4}$ M$_{\sun}$ (left), and all clusters with masses between 500 - 10000 M$_{\sun}$ (right).  What we have found is that all clusters with ages that lie between 6-8 Myr have non-detections in H$\alpha$. All clusters younger than 4.5 Myr are detected in H$\alpha$, and 2/3 of clusters between 4.5 and 6 Myr are detected in H$\alpha$. We must note that the uncertainties in each bin of Figure 4 are larger than the bin size, yet even when accounting for the uncertainties in the best fit age, the ages of clusters showing H$\alpha$ remain below 6 Myr. The exception here are those few clusters with masses $>$ 10$^{4}$ M$_{\sun}$.  These have not been included in  Figure \ref{fig:agewithha} because even at ages between 2-3 Myr, they have very little measured H$\alpha$ flux, likely due to stronger feedback effects from the clusters themselves, or the more crowded environment in which they are located which makes it difficult to determine which ionizing photons come from which cluster. \citet{2012MNRAS.423.2933R} have also determined that the leakage of ionizing photons is expected for those HII regions with ages greater than 4 Myr. While some of the detected clusters do venture into ages greater than 5 Myr, we do need to be aware of the fact that especially in clusters that may contain only one or two extremely massive stars they may not live long enough to produce ionizing photons out to 8 Myr.  For example, the lifetime of a 35 M$_{\sun}$ star is roughly 5 Myr, while a 15 M$_{\sun}$ star may live 100 Myr and a 120 M$_{\sun}$ star only 2 Myr.  If there is only one massive star in these smaller clusters, the L$_{H\alpha}$  may be more sensitive to the age.
\begin{figure} 
   \centering 
      \includegraphics[width=2.5in, angle=270]{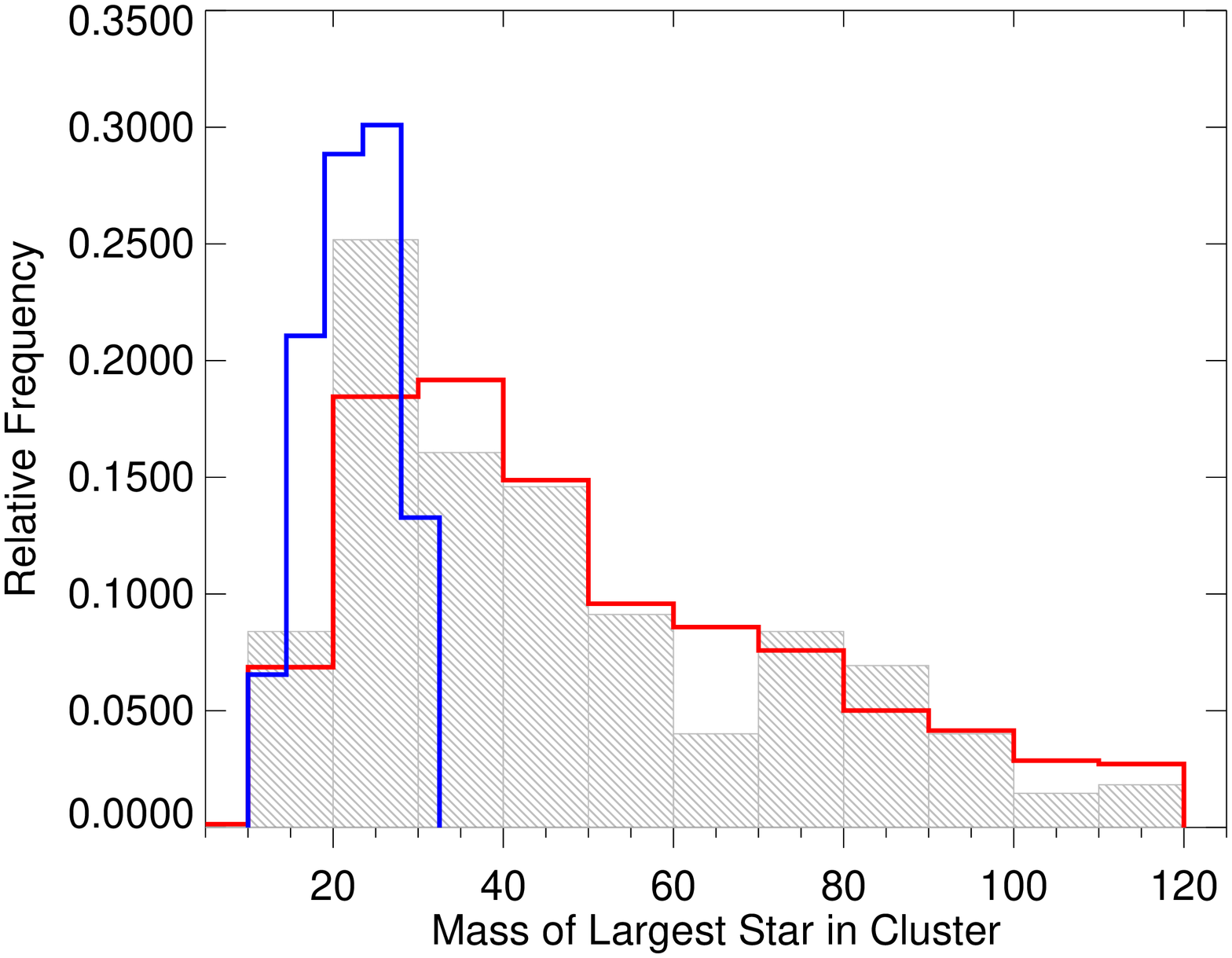} 
   \caption{Histogram of the most massive star in each model considered the best fit along with fits within $\Delta \chi ^{2}$ of 10 $\%$ of the lowest value from SLUG plotted against a fully sampled IMF (red) and a model truncated at 30 M$_{\sun}$ (blue). Masses range from 10-119 M$_{\sun}$, and while 50$\%$ of the most massive stars are $<$ 40 M$_{\sun}$, there are still the existence of stars between 40-120 M$_{\sun}$ in clusters with masses of 10$^{3}$ M$_{\sun}$ which would be contrary to a variable upper mass limit \citep{2003ApJ...598.1076K,2010MNRAS.401..275W} theory.}
   \label{fig:mostmass}
\end{figure}

\begin{figure*}
   \centering
   \includegraphics[width=2.5in,angle=270]{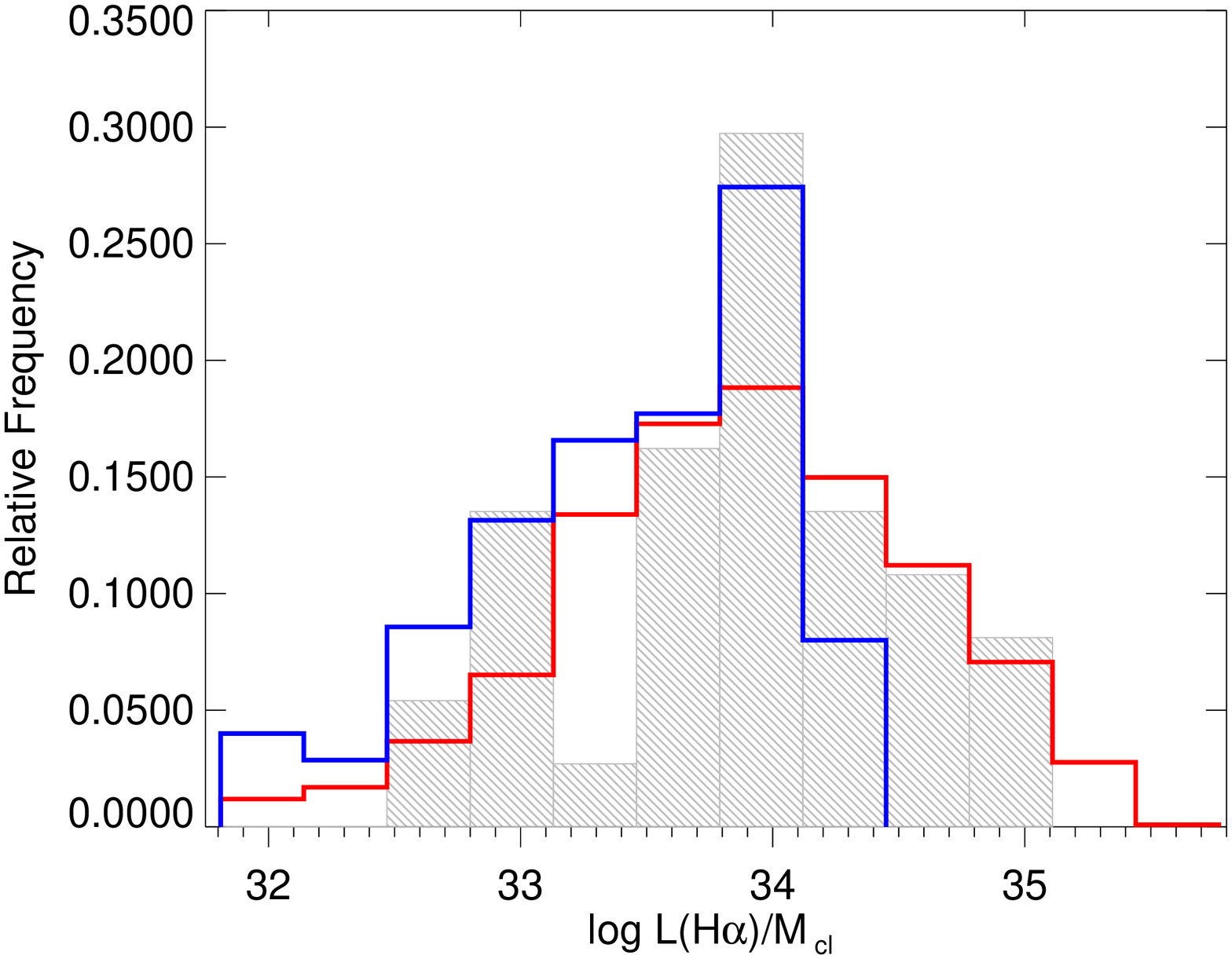} 
    \includegraphics[width=2.5in,angle=270]{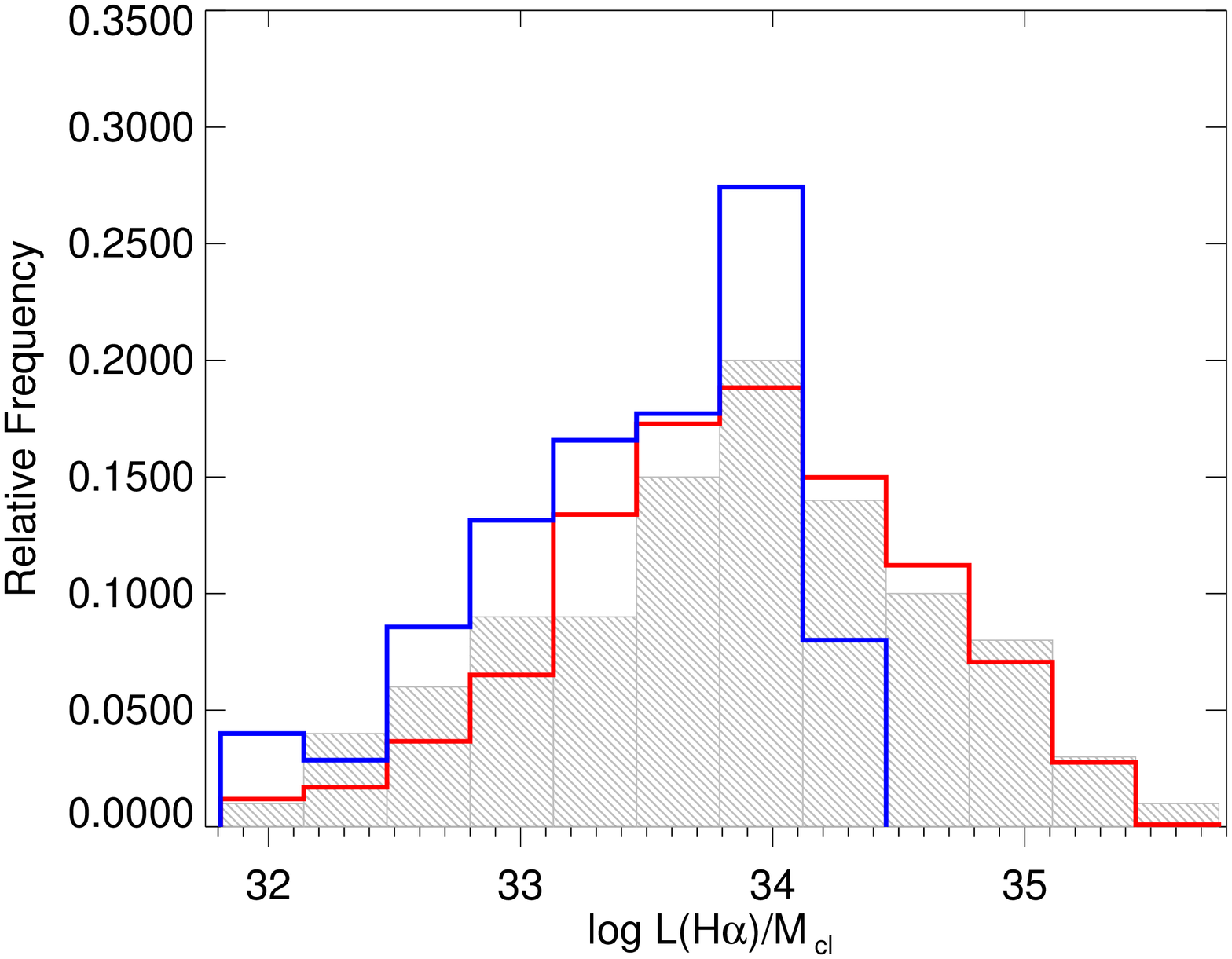}
   \includegraphics[width=2.5in,angle=270]{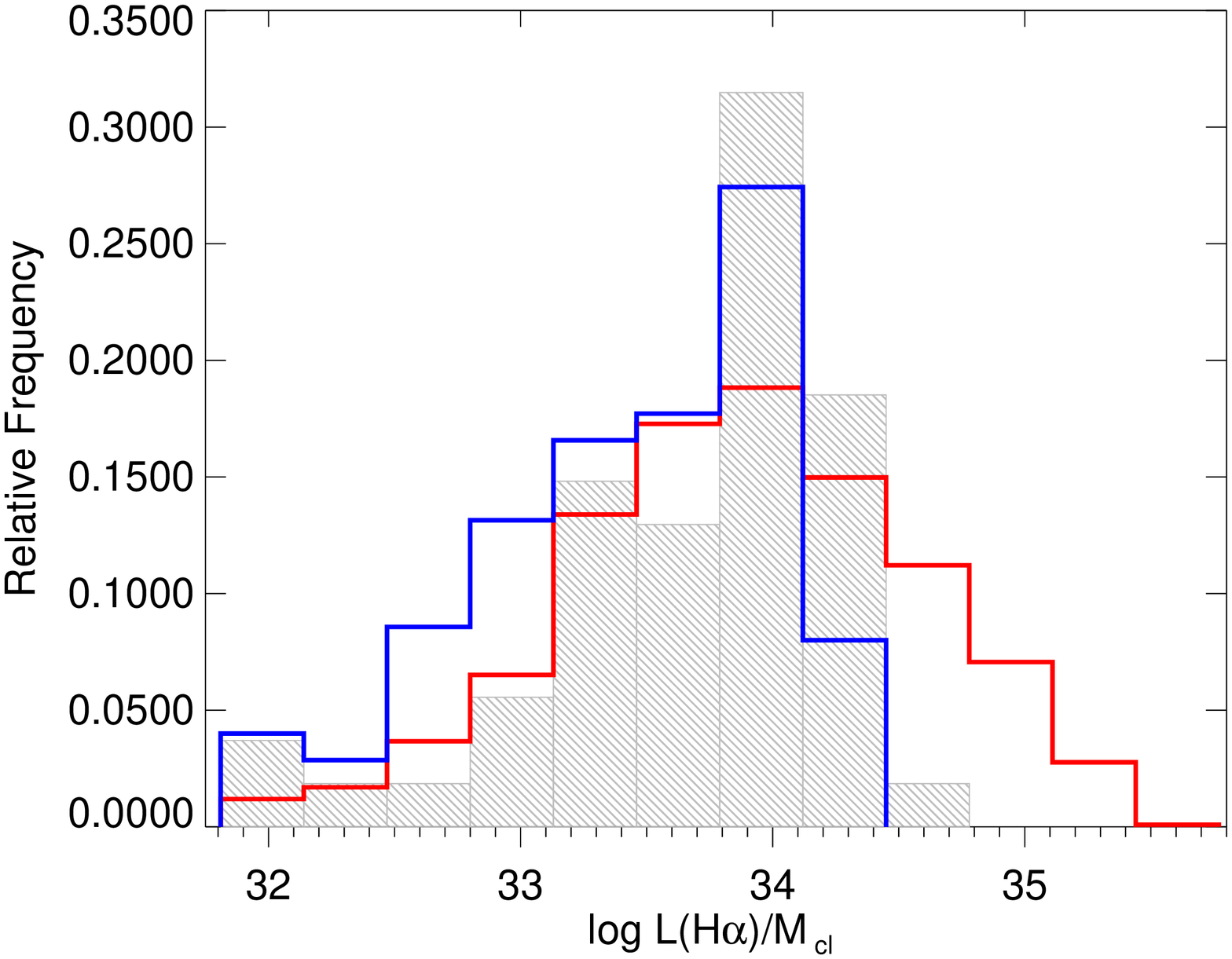}
    \includegraphics[width=2.5in,angle=270]{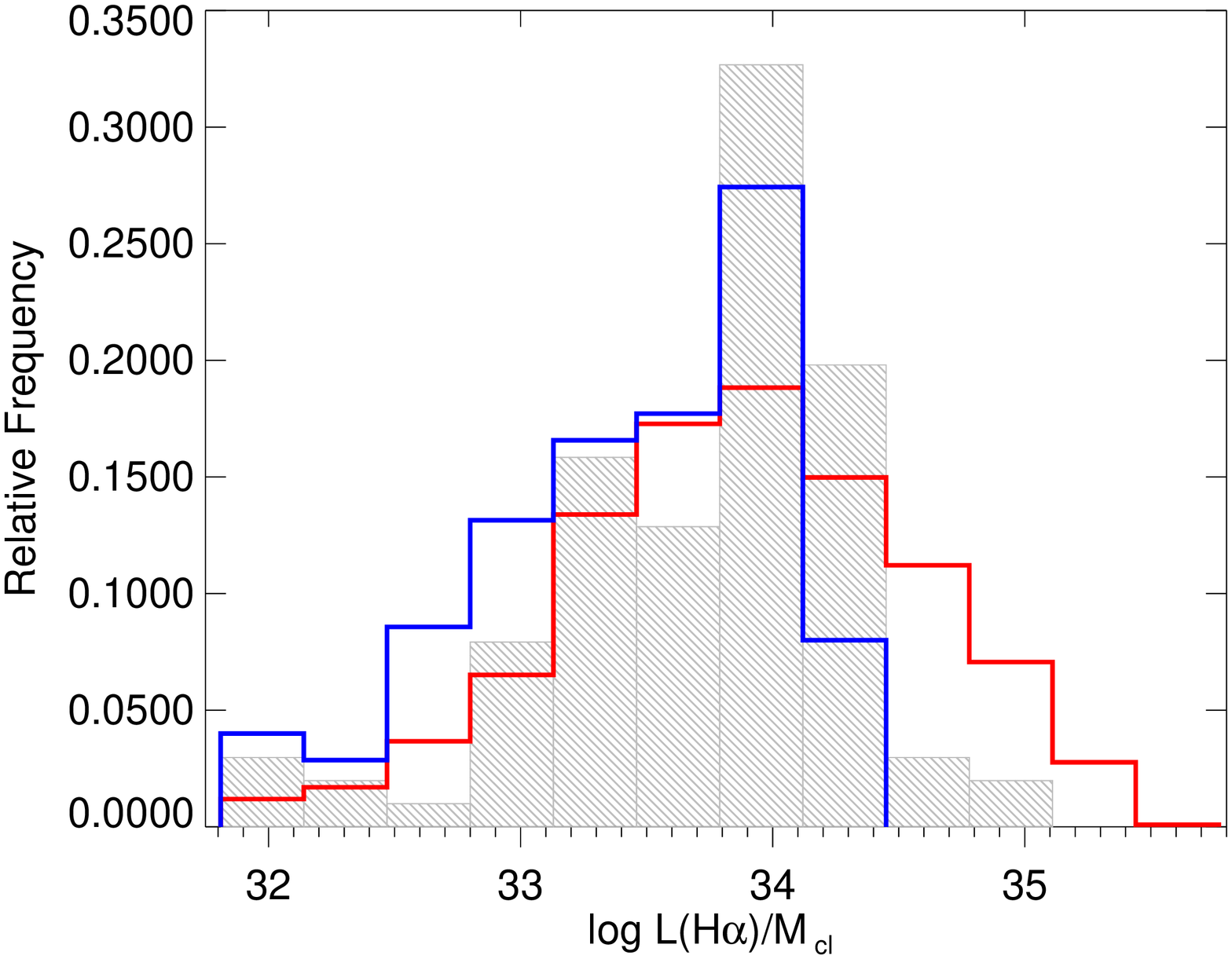}
   \caption{Histograms of L$_{H\alpha}$ /M$_{cl}$ from SLUG models for a fully sampled IMF (red) and a model truncated at 30 M$_{\sun}$ (blue) for L$_{H\alpha}$ /M$_{cl}$ plotted against clusters from NGC 4214 whose masses were determined from SLUG models with a maximum stellar mass of 120 M$_{\sun}$ (top) and SB99 models with a maximum stellar mass of 30 M$_{\sun}$ (bottom). The left panels only include the single best-fit mass, while the right panels include error bars for all masses with a $\chi ^{2}$ $<$ 1.}
   \label{fig:histo}
\end{figure*}

For each model, we show a histogram (Figure 6) of the most massive star in the model which best describes the photometry in each of the 47 clusters located in the two least mass bins. Included in this plot are also the error bars associated with the ensemble of fits within $\Delta \chi ^{2}$ of 10$\%$ of the lowest value, and model lines for a fully sampled IMF (red) and a model truncated at 30 M$_{\sun}$ (blue). What we find is that 25$\%$ of these model clusters have a maximum mass between 20-30 M$_{\sun}$, 50$\%$ have a mass $<$ 40 M$_{\sun}$, and the remaining 50$\%$ contain most massive stars greater than 40 M$_{\sun}$. We must point out that because clusters less than 500 M$_{\sun}$ were excluded from the study it is possible that we may be underestimating the most massive stars with masses $<$ 20 M$_{\sun}$.  According to \citet{2011arXiv1112.3340K}, specifically Figures 2 and 5, the expected maximum stellar mass of a cluster between 500-3000 M$_{\sun}$ is between 20-40 M$_{\sun}$ if the \citet{2010MNRAS.401..275W} hypothesis holds. Some of our clusters have a maximum stellar mass at this cluster mass in Figure \ref{fig:mostmass}, but the existence of stellar masses greater than 40 M$_{\sun}$ for the same range of cluster masses indicates that there is no maximum mass for the stars in these clusters, other than the usual upper limit found in much more massive clusters. Our results in Figure \ref{fig:mostmass} are consistent with the random sampling from Figure 2 of \citet{2011arXiv1112.3340K}, where the maximum stellar mass can range between 15-120 M$_{\sun}$ for a 500-3000 M$_{\sun}$ cluster. We must point out that these models were populated specifically for 1 $\times$ 10$^{3}$ M$_{\sun}$ clusters, and the clusters themselves range between 500-9000 M$_{\sun}$, but that our treatment should give a reasonable representation of the actual spread of massive stars. As mentioned above, tests run on 500 M$_{\sun}$, 1000 M$_{\sun}$, and 3000 M$_{\sun}$ SLUG models resulted in mass and age discrepancies within the uncertainties of the $\chi ^{2} < $ 1 results.

\citet{2002AJ....124.1393L} has suggested that lower SFR correlates with a lower average cluster mass; therefore most galaxies with a low SFR should be deficient in high mass clusters in a purely stochastic case. For example, 100 galaxies with a SFR of 0.1 M$_{\sun}$ yr$^{-1}$ will have the same distribution of cluster masses as 1 galaxy with a SFR of 10 M$_{\sun}$ yr$^{-1}$, yet most of the low SFR galaxies will be lacking massive clusters and a few will have an over abundance.  Other studies have indicated that there is a systematic trend for small clusters to form only low mass stars \citep[for example]{2003ApJ...598.1076K,2010MNRAS.401..275W,2009MNRAS.395..394P}, effectively steepening the the slope of the uIMF \citep{2009ApJ...695..765M}.  When interpolated across a galaxy, the integrated galactic IMF \citep[IGIMF]{2003ApJ...598.1076K,2010MNRAS.401..275W} would ultimately lead to low SFR galaxies such as dwarfs to be deficient in both large stellar clusters and massive stars.  From what we have presented above, we do not find that this is the case in NGC 4214.  The SLUG models suggest that roughly 50$\%$ of clusters with masses around 1 $\times$ 10$^{3}$ M$_{\sun}$ have at least one star more massive than 40 M$_{\sun}$ (Figure \ref{fig:mostmass}).  We have also performed a Komogorov-Smirnov (K-S) test between the best-fit L$_{H\alpha}$ /M$_{cl}$ data and both SLUG models sampled up to 30 M$_{\sun}$ and 120 M$_{\sun}$ (blue and red lines in Figure \ref{fig:histo}, respectively). We find that there is about a 65$\%$ chance the data were drawn from the parent model of the fully populated Kroupa IMF up to 120 M$_{\sun}$, but only about a 1$\%$ chance they come from the models truncated at 30 M$_{\sun}$. By adding in error bars from the ensemble of fits within $\Delta \chi ^{2}$ (right), these numbers become 75$\%$ and 1$\%$ respectively . When we compare the  L$_{H\alpha}$ /M$_{cl}$ values of the clusters which were age-dated using a SB99 model where the most massive star is only allowed to be 30 M$_{\sun}$ against the fully populated and truncated SLUG models (see Figure \ref{fig:histo}, bottom) the K-S test only shows agreement of 7$\%$ and 4$\%$ respectively that the data come from those respective populations, even when uncertainties are included.  Thus, even allowing our clusters to be modeled as drawn from a parent population that does not have stars more massive than 30 M$_{\sun}$ (which affect the determination of the cluster masses) the L$_{H\alpha}$ /M$_{cl}$ distribution does not change in such a way to agree with a truncated IMF.

\section{Conclusions}
We have attempted to constrain the upper end of the IMF of the nearby irregular star bursting dwarf galaxy NGC 4214 using the methods of  \citet{2010ApJ...719L.158C}, which uses the ratio of the luminosity of the ionizing photons normalized to the mass of the cluster as a proxy for probing the presence of massive stars.  With a final sample of 52  young clusters with masses $>$ 500 M$_{\sun}$, we have determined that even at masses $\sim$ 10$^{3}$ M$_{\sun}$, there does not seem to be a deviation from the expected ionizing flux of an universal IMF up to 120 M$_{\sun}$.  Clusters with a mean mass down to 700 M$_{\sun}$ have a L$_{H\alpha}$ /M$_{cl}$ ratio that lies along that predicted by a universal IMF.

We have also determined that a truncated IMF (one in which the maximum M$_{*}$ is a function of M$_{cl}$), which would result in the maximum stellar mass in a cluster of 10$^{3}$ M$_{\sun}$ being no greater than 35 M$_{\sun}$ \citep{2010MNRAS.401..275W}, does not sufficiently explain the young clusters seen in NGC 4214.  Models used to age-date the clusters indicate that up to 50$\%$ of the clusters contain a massive star greater than 40 M$_{\sun}$, while Komogorov-Smirnov tests indicate that there is only a 1$\%$ chance that the L$_{H\alpha}$ /M$_{cl}$ values from clusters in NGC 4214 come from a parent model with the maximum mass truncated at 30 M$_{\sun}$ and a 75$\%$ chance that they arise from a fully populated IMF up to 120 M$_{\sun}$.  As with \citet{2011ApJ...741L..26F}, who investigated the integrated properties of individual galaxies, we find that the summation of individual young clusters in NGC 4214 is more consistent with a universal IMF without a truncation of massive stars. Our test shows not only that the M$_{*}$ -- M$_{cl}$ relation of \citet{2010MNRAS.401..275W} is in disagreement with the observations, but also that the data are consistent with a stochastically-sampled IMF with our upper mass limit consistent with a standard \citet{2001MNRAS.322..231K} IMF.

\acknowledgements{We would like to thank the referee for valuable comments on the paper.  J.A. and D.C. acknowledge partial support for this study from the grant associated with program \# GO-11360 (P.I.: R.W. O'Connell), which was provided by NASA through the Space Telescope Science Institute, which is operated by the Association of Universities for Research in Astronomy, Inc., under NASA contract NAS 5-26555. MRK acknowledges support from an Alfred P. Sloan Fellowship, from the NSF through grant CAREER-0955300, and from NASA through a Chandra Space Telescope Grant and through Astrophysics Theory and Fundamental Physics Grant NNX09AK31G. The work of R.L.dS. is supported by a National Science Foundation Graduate Research Fellowship.}

\appendix
\section{Filter Convolution}
SLUG models, as of the submission of this paper, are not delivered in HST filter pass bands. Luminosities were delivered in Johnson-Cousins UBVI and Galex NUV, and, because of this, transformation coefficients were derived to approximate WFC3 filters.  To derive these coefficients, each 1 Myr time step of the SB99 model was convolved with the appropriate WFC3 filter passband (discussed below) to get the fluxes in each filter at each age.  The SLUG models were then divided into 1 Myr age bins (for example, the 3 Myr age bin consisted of all models between the ages of 2.5 and 3.4 Myr), and the flux in each filter was averaged into a mean flux.  Therefore for  1 Myr age bins, there exists an average luminosity for each filter to directly compare with SB99 models.  The ratio between the SB99 and SLUG models was then taken, resulting in coefficients to convert SLUG filters to WFC3 filters.  These values were then applied back onto each of the individual $\sim$ 40000 SLUG models, and the effective wavelengths were adjusted to reflect those of WFC3 filters. 

For the SB99 models, the filter convolution process was more straightforward.  From the output SED each 1 Myr age bin was extracted and then using the IRAF function $sinterp$ were all interpolated into the same wavelength range of 500-10000 \AA.  Each WFC3/UVIS filter transmission curve was also extracted into the identical wavelength range, and then integrated over those wavelengths to determine the total transmission for each filter at the effective wavelength ($\int T_{\lambda}d\lambda$).  The SB99 SEDs were then multiplied by the transmission curves and a numerical integrated value was obtained ($\int F_{\lambda}T_{\lambda}d\lambda$).  Next, $<F_{\lambda}>$ = $\frac{\int F_{\lambda}T_{\lambda}d\lambda}{\int T_{\lambda}d\lambda}$ is then calculated for each age bin and each filter.  This transforms the continuous SB99 SED into 5 distinct photometry points which can be easily compared with the photometry of the NGC 4214 clusters (Figure 3).

Once the models have been convolved with the filters, they are then corrected for possible host galaxy extinction. Due to the low metallically of this galaxy, we have chosen to use an SMC extinction curve from \citet{1999PASP..111...63F} which is more in line with the metallically of NGC 4214.  We have limited our reddening to lie between 0.0 $\leq$ $E(B-V)$ $\leq$ 0.40 as previous studies on this galaxy seem to indicate low values of extinction \citep{2007AJ....133..932U}. The population synthesis SEDs are first convolved with the extinction curve at selected values of the color excess \textit{E(B-V),
spanning the full range of \textit{E(B-V)} used for our analysis (0.0-0.4), in order to determine the effective wavelengths $\lambda_{ext,eff}$ to be used when applying extinction to broad band filter photometry. In general $\lambda_{ext,eff}$ is different from the effective wavelength of a filter, owing to the non-symmetric transmission curves of most filters. We should note that $\lambda_{ext,eff}$ depends also on the range of E(B-V) considered, and should be re-calculated when using color excesses outside our range.} We elect to apply extinction corrections after filter convolution because tests comparing the results of this procedure against the other procedure of applying the extinction before filter convolution gives differences of only 5$\%$.  Our choice, however, provides the flexibility to change $E(B-V)$ at will.  Each filter is then multiplied by $e ^{0.92 \times E(B-V) \times SMC}$, where $E(B-V)$ is in increments of 0.02, and the appropriate SMC extinction value is determined for the wavelength of each filter.

\clearpage
\renewcommand{\thefootnote}{\alph{footnote}}
\small
\begin{longtable}{ccccccccccc}
\caption{Ages, Masses, and Extinctions of Clusters}\\
\hline \hline \\
& && & \underline{SLUG}& && \underline{SB99}&&& \\
ID&$\alpha$ (J2000) &$\delta$ (J2000) & Age& Mass& $E(B-V)$&Age & Mass & $E(B-V)$& L$_{H\alpha}$\tablenotemark{a}&Other\\
& 12$^{h}$15$^{m}$$+$&36$^{\circ}$$+$& (Myr)& (10$^{3}$ M$_{\sun}$)& &(Myr)&  (10$^{3}$ M$_{\sun}$)&& erg s$^{-1}$&Names\tablenotemark{b}\\
\hline\\
\endfirsthead
\hline \hline \\
& && & \underline{SLUG}& && \underline{SB99}&&& \\
ID&$\alpha$ (J2000) &$\delta$ (J2000) & Age& Mass& $E(B-V)$&Age & Mass & $E(B-V)$& L$_{H\alpha}$&Other\\
& 12$^{h}$15$^{m}$$+$&36$^{\circ}$$+$& (Myr)& (10$^{3}$ M$_{\sun}$)& &(Myr)&  (10$^{3}$ M$_{\sun}$)&& erg s$^{-1}$&Names\tablenotemark{b}\\
\hline \\
\endhead \hline
\endfoot
\\ \endlastfoot
55\tablenotemark{c}	&	39.278	&	19 55.94	&	5.9$\pm0.9$	&	0.50$^{0.2}_{0.6}$	&	0.24	&	4.0	&	0.21	&	0.16	&	1.6E+35&	\\
145$^{c}$	&	40.065	&	19 14.13	&	4.6$\pm3.0$	&	0.54$^{0.5}_{1.2}$	&	0.16	&	5.0	&	0.12	&	0.2	&	1.6E+35&	\\
115	&	37.723	&	19 46.42	&	6.6$\pm2.0$&	0.57$^{0.4}_{2.5}$	&	0.32	&	4.0	&	0.50	&	0.4	&	1.6E+35&	\\
68$^{c}$	&	37.008	&	19 54.39	&	4.7$\pm3.0$	&	0.60$^{0.6}_{3.7}$	&	0.22	&	5.0	&	0.33	&	0.22	&	1.6E+35&	\\
63	&	40.559	&	19 24.44	&	5.6$\pm2.3$	&	0.70$^{0.7}_{2.0}$	&	0.22	&	5.4	&	0.20	&	0.12	&	1.6E+35&	\\
159$^{c}$	&	39.072	&	19 38.38	&	4.7$\pm3.0$	&	0.86$^{0.9}_{2.5}$	&	0.3	&	5.6	&	0.20	&	0.26	&	1.6E+35&	\\
188	&	37.928	&	21 05.26	&	7.5$\pm0.8$	&	2.10$^{2.0}_{2.1}$	&	0.4	&	7.6	&	0.40	&	0.34	&	1.6E+35&	\\
27	&	38.508	&	19 45.84	&	5.9$\pm1.3$	&	2.22$^{1.3}_{5.6}$	&	0.32	&	4.0	&	1.00	&	0.26	&	1.6E+35	&\\
17	&	40.186	&	21 03.80	&	5.2$\pm1.7$	&	2.44$^{2.4}_{3.4}$	&	0.06	&	5.8	&	0.49	&	0.02	&	1.6E+35	&\\
34	&	38.653	&	20 00.20	&	5.4$\pm1.9$	&	2.73$^{2.7}_{1.1}$	&	0.12	&	5.4	&	0.46	&	0.1	&	1.6E+35&	\\
2	&	40.372	&	19 29.87	&	3.1$\pm0.5$	&	13.04$^{0.1}_{2.7}$	&	0.06	&	4.8	&	3.60	&	0.04	&	1.6E+35&	I-B2n\\
3	&	40.487	&	19 31.71	&	2.6$\pm0.5$	&	20.03$^{2.7}_{0.9}$	&	0.08	&	4.8	&	4.58	&	0.08	&	1.6E+35&I-Bs	\\
4	&	40.926	&	19 27.09	&	2.9$\pm1.5$	&	20.71$^{5.0}_{1.9}$	&	0.06	&	5.2	&	3.77	&	0.06	&	1.6E+35&	\\
\hline
140	&	40.925	&	18 54.40	&	5.9$\pm2.7$	&	0.49$^{0.4}_{2.1}$	&	0.38	&	4.0	&	0.21	&	0.3	&	2.3E+36&	\\
340	&	40.555	&	19 12.84	&	3.7$\pm1.7$	&	0.52$^{0.5}_{2.0}$	&	0.16	&	3.8	&	0.38	&	0.16	&	3.1E+37& II-C1n	\\
348$^{c}$	&	38.214	&	18 40.35	&	3.1$\pm1.7$	&	0.57$^{0.5}_{0.4}$	&	0	&	4.6	&	0.17	&	0	&	3.7E+36&	\\
359$^{c}$	&	39.226	&	19 41.53	&	3.6$\pm1.5$	&	0.57$^{0.3}_{6.9}$	&	0.2	&	3.2	&	1.01	&	0.22	&	8.8E+36&	\\
345	&	40.985	&	18 55.60	&	2.7$\pm2.4$	&	0.61$^{0.6}_{3.5}$	&	0.28	&	4.2	&	0.39	&	0.28	&	4.0E+36&	\\
74$^{c}$	&	38.679	&	19 44.18	&	5.6$\pm2.7$	&	0.65$^{0.6}_{2.0}$	&	0.24	&	5.2	&	0.19	&	0.16	&	2.3E+36&	\\
25	&	39.566	&	19 32.93	&	3.5$\pm1.2$	&	0.66$^{0.5}_{0.7}$	&	0.14	&	3.2	&	0.39	&	0.14	&	5.1E+37&	I-A5n\\
402$^{c}$	&	38.659	&	19 31.17	&	2.2$\pm1.5$	&	0.69$^{0.6}_{0.9}$	&	0.06	&	3.6	&	0.22	&	0.08	&	5.1E+36&I-C1n	\\
46	&	40.908	&	19 23.26	&	2.9$\pm1.9$	&	0.72$^{0.7}_{0.5}$	&	0	&	5.2	&	0.13	&	0	&	7.0E+35&	\\
125$^{c}$	&	45.614	&	19 17.81	&	5.6$\pm3.0$	&	0.76$^{0.8}_{0.8}$	&	0.22	&	5.4	&	0.10	&	0.14	&	1.1E+37&	\\
58	&	39.713	&	19 25.88	&	4.3$\pm2.7$	&	0.77$^{0.7}_{1.1}$	&	0.04	&	4.8	&	0.13	&	0.04	&	9.7E+35&	\\
24	&	39.613	&	19 34.57	&	4.9$\pm1.8$	&	0.78$^{0.7}_{5.8}$	&	0.18	&	5.2	&	0.86	&	0.14	&	7.3E+36&	\\
417	&	41.921	&	19 12.33	&	5.9$\pm3.0$	&	0.80$^{0.7}_{7.3}$	&	0.22	&	5.0	&	0.62	&	0.14	&	3.9E+36&	\\
328	&	34.729	&	20 17.91	&	5.9$\pm2.8$	&	0.85$^{0.8}_{4.7}$	&	0.2	&	4.0	&	0.35	&	0.12	&	2.8E+37&	\\
39	&	43.649	&	19 00.15	&	2.9$\pm1.6$	&	0.88$^{0.8}_{0.4}$	&	0	&	5.4	&	0.17	&	0	&	3.6E+36&	\\
384	&	39.067	&	19 45.63	&	5.9$\pm1.5$	&	0.93$^{0.8}_{1.3}$	&	0.24	&	4.0	&	0.38	&	0.16	&	1.1E+37&	\\
21	&	41.303	&	20 29.18	&	4.7$\pm1.5$	&	0.94$^{0.8}_{2.6}$	&	0.04	&	5.0	&	0.34	&	0.02	&	9.5E+36&	\\
360	&	38.996	&	19 37.41	&	4.5$\pm2.0$	&	0.95$^{0.7}_{7.8}$	&	0.18	&	4.0	&	0.95	&	0.16	&	7.5E+37&I-A3n	\\
326	&	42.110	&	19 01.12	&	4.0$\pm1.5$	&	1.14$^{0.9}_{3.6}$	&	0.4	&	4.0	&	0.77	&	0.38	&	3.9E+37&IXn	\\
83	&	41.017	&	19 01.41	&	5.9$\pm2.8$	&	1.18$^{1.0}_{5.2}$	&	0.4	&	5.0	&	0.58	&	0.32	&	2.2E+37&	II-A\\
358	&	39.285	&	19 47.16	&	5.1$\pm1.5$	&	1.26$^{1.0}_{4.1}$	&	0.12	&	5.0	&	0.85	&	0.1	&	2.5E+37&I-D2n	\\
352	&	38.482	&	18 45.42	&	3.1$\pm0.5$	&	1.31$^{1.2}_{0.4}$	&	0	&	4.6	&	0.39	&	0	&	2.1E+37&	\\
341	&	40.611	&	19 12.53	&	3.5$\pm1.5$	&	1.35$^{1.2}_{4.3}$	&	0.14	&	3.2	&	0.78	&	0.14	&	3.9E+37&	II-C1n\\
362	&	38.809	&	19 31.27	&	3.4$\pm1.2$	&	1.53$^{0.5}_{1.7}$	&	0.12	&	4.4	&	3.39	&	0.08	&	1.6E+37&	I-C1n\\
357	&	34.527	&	19 46.39	&	5.6$\pm2.0$	&	1.58$^{0.4}_{5.0}$	&	0.08	&	5.0	&	0.57	&	0	&	2.5E+36&	\\
338	&	40.700	&	19 09.83	&	6.0$\pm0.4$	&	1.62$^{0.1}_{1.0}$	&	0	&	6.2	&	5.75	&	0.08	&	1.0E+38&	II-B\\
81	&	39.087	&	19 39.71	&	4.7$\pm2.2$	&	1.63$^{1.3}_{6.0}$	&	0.4	&	5.0	&	0.90	&	0.4	&	8.6E+35&	\\
434$^{c}$	&	36.612	&	20 06.30	&	3.4$\pm1.7$	&	1.77$^{1.3}_{1.6}$&	0.38	&	4.8	&	0.43	&	0.1	&	2.2E+36&	\\
15	&	40.119	&	19 26.67	&	3.4$\pm1.5$	&	2.16$^{2.0}_{0.8}$	&	0.02	&	4.8	&	0.57	&	0.04	&	1.6E+36&	\\
349	&	38.358	&	18 47.05	&	2.5$\pm1.5$	&	2.66$^{2.2}_{0.4}$&	0	&	4.8	&	0.56	&	0.22	&	1.2E+37&	\\
100	&	40.398	&	18 51.19	&	5.5$\pm2.8$	&	2.67$^{2.6}_{2.1}$	&	0.28	&	5.8	&	0.26	&	0.2	&	7.9E+35&	\\
353	&	37.617	&	19 00.90	&	3.3$\pm2.2$	&	3.08$^{3.0}_{5.5}$&	0.28	&	4.6	&	1.20	&	0.26	&	1.9E+37&	\\
11	&	40.384	&	19 30.80	&	5.4$\pm0.7$	&	3.39$^{2.8}_{3.2}$	&	0.2	&	5.6	&	4.29	&	0.16	&	2.8E+36&	I-B2n\\
9	&	40.662	&	19 14.11	&	2.7$\pm1.2$	&	3.70$^{3.3}_{2.5}$	&	0	&	4.6	&	1.24	&	0.02	&	1.9E+37&	 II-C2n\\
18	&	41.076	&	19 29.83	&	4.2$\pm1.0$	&	3.92$^{3.5}_{2.1}$	&	0.2	&	4.8	&	1.31	&	0.2	&	4.8E+37& I-Gn	\\
365	&	40.694	&	19 09.93	&	4.1$\pm0.8$	&	5.88$^{4.5}_{2.6}$&	0.18	&	4.0	&	3.22	&	0.14	&	5.1E+37& II-B	\\
6	&	39.250	&	19 33.95	&	4.4$\pm0.8$	&	8.21$^{7.1}_{7.1}$	&	0.14	&	4.0	&	2.68	&	0.12	&	9.7E+37&	I-A1n\\
395	&	39.186	&	19 30.32	&	2.1$\pm0.7$	&	35.01$^{3.1}_{2.9}$	&	0.12	&	4.6	&	7.03	&	0.14	&	9.0E+36&	\\
1	&	39.442	&	19 34.94	&	4.2$\pm1.6$	&	94.00$^{40.0}_{40.0}$	&	0.14	&	4.8	&	97.81	&	0.1	&	8.4E+37\tablenotemark{d}& I-As	\\
\hline
\centering

\footnotetext[1]{Clusters above the line have only 3$\sigma$ detections of H$\alpha$.  Clusters 1-4 likely contain H$\alpha$ emission, but due to size and nearness of other clusters have blown cavities surrounding them.}
\footnotetext[2]{Other names come from MacKenty et al. 2000.}
\footnotetext[3]{PSF consistent with a single star at the distance of NGC 4214 (FWHM $\sim$ 1.3 pc), and SED colors consistent with either a single O star with several low mass (B5 or later) stars or a few early B type stars.}
\footnotetext[4]{Luminosity derived from Table 3 in \citet{2000AJ....120.3007M}.}
\label{tab:resultstable}
\end{longtable}

\end{document}